\documentclass{aa}

\usepackage{amsmath,amssymb}

\usepackage{bm}
\usepackage{color}
\usepackage{float}
\usepackage{graphicx}
\usepackage{txfonts}
\usepackage{natbib}
\usepackage[normalem]{ulem}
\def\MODHOL#1{{\textcolor{black}{#1}}}

\def\cwoadd#1{{\textcolor{black}{#1}}}
\newcommand{\cwosug}[2]{{\textcolor{black}{#2}}}
\newcommand{\JB}[1]{\textcolor{black}{#1}} 
\newcommand{\TG}[1]{\textcolor{black}{#1}} 

\def\wig#1{\mathrel{\hbox{\hbox to 0pt{%
          \lower.6ex\hbox{$\sim$}\hss}\raise.4ex\hbox{$#1$}}}}

\def\lkol{\ell_{\rm Kol}}
\def\Kol{{\rm Kol}}
\def\nut{\nu_{\rm t}}
\def\v{{\rm v}}   
\def\Rep{{Re_{\rm p}}}
\def\taus{t_{\rm s}}  
\def\tauc{t_{\rm c}}  
\newcommand{\St}{\mbox{\textit{St}}}

\begin{document}

   \title{Effect of turbulence on collisions of dust particles with
     planetesimals in protoplanetary disks}

   \author{H. Homann
          \inst{1}
          \and
          T. Guillot
          \inst{1}
          \and
          J. Bec
          \inst{1}
          \and
          C. W. Ormel
          \inst{2}
          \and
          S. Ida
          \inst{3,4}
          \and
          P. Tanga
          \inst{1}
          }

   \institute{Laboratoire J.-L.\ Lagrange, Universit\'e C\^ote d'Azur, Observatoire de la
     C\^ote d'Azur, CNRS, F-06304 Nice, France\\
     \email{holger.homann@oca.eu}
     \and
     Anton Pannekoek Institute for Astronomy, University of Amsterdam, The Netherlands
     \and
     Department of Earth and Planetary Sciences, Tokyo Institute of Technology, 152-8551 Tokyo, Japan
     \and
     Earth-Life Science Institute, Tokyo Institute of Technology, 152-8550 Tokyo, Japan\\
   }

   \date{DRAFT: Do not distribute, \today}

 
  \abstract
   {Planetesimals in gaseous protoplanetary disks may grow by
     collecting dust particles. Hydrodynamical studies show that small
     particles generally avoid collisions with the planetesimals
     because they are entrained by the flow around them. This occurs
     when $\St$, the Stokes number, defined as the ratio of the dust
     stopping time to the planetesimal crossing time, becomes much
     smaller than unity. However, these studies have been limited to
     the laminar case, whereas these disks are believed to be
     turbulent.}
   {We want to estimate the influence of gas turbulence on the
     dust-planetesimal collision rate and on the impact speeds.}
   {We used three-dimensional direct numerical simulations of a fixed
     sphere (planetesimal) facing a laminar and turbulent flow seeded
     with small inertial particles (dust) subject to a Stokes drag. A
     no-slip boundary condition on the planetesimal surface is modeled
     via a penalty method.}
   {We find that turbulence can significantly increase the collision
     rate of dust particles with planetesimals. For a high turbulence
     case (when the amplitude of turbulent fluctuations is similar to
     the headwind velocity), we find that the collision probability
     remains equal to the geometrical rate or even higher for
     $\St\wig{>}0.1$\TG{, i.e., for dust sizes an order of magnitude
       smaller than in the laminar case}.  We derive expressions to
     calculate impact probabilities as a function of dust and
     planetesimal size and turbulent intensity. }
   {}

   \keywords{Planets and satellites: formation, Planet-disk
     interactions, Turbulence, Accretion, Accretion disks, hydrodynamics, methods: numerical }

   \maketitle
%

\section{Introduction}

Conventional models for planet formation involve the hierarchical
growth by accretion of ``planetesimals''.  Such building blocks
undergo collisions and gravitational binding to eventually reach
planetary sizes \citep[see,
  e.g.,][]{Safronov1972,hayashi1985formation}. Still several serious
uncertainties remain in the processes leading from sub-$\mu$m dust
grains, which follow the sub-Keplerian gas, to km-sized planetesimals
that are massive enough to move on Keplerian orbits.  One of them is
known as the ``meter-size barrier''. \cwosug{Bodies with sizes of the
  order of the meter have a strong drag with the gas that dissipates
  angular momentum and thus undergo a very rapid orbital
  decay}{Meter-sized bodies experience a strong drag force, which
  causes rapid orbital decay due to angular momentum loss}. The
timescale of this decay is less than 100 years at 1AU \citep[see,
  e.g.,][]{weidenschilling1984evolution,nakagawa1986settling}, which
is shorter by several orders of magnitude than the observationally
inferred disk lifetime (several million years). \cwosug{[I suggest to
    add:]}{Consequently, the planet-forming material are lost from the
  disk.}

\cwoadd{One scenario for overcoming the meter-size barrier is by the
  combined effect of streaming instabilities and pebble
  accretion. Because of growth to mm/cm-sized particles and settling,
  dust concentrates in the midplane and may become unstable to
  streaming instabilities \citep{Youdin+Goodman2005, Johansen+2007,
    johansen2011high}. This forms relatively large
  planetesimals. After that, these large planetesimals can sweep up
  surrounding grains and migrating pebbles
  \citep{ormel2010effect,ormel2012understanding,lambrechts2012rapid}.}
Because the orbital decay of the pebbles is also fast, a large number
of pebbles are supplied from outer disk regions on relatively short
timescales, resulting in rapid planet growth.  This scenario utilizes
the too rapid migration problem, rather than avoids it.  The pebble
accretion model is actively discussed for the formation of the solar
system \citep{morbidelli2015great} and of close-in super-Earth systems
in exoplanetary systems
\citep{chatterjee2014inside,chatterjee2015vulcan}.  However, the
details of the \cwosug{formation of pebbles and of}{} the accretion of
grains \cwoadd{and pebbles} by (large) planetesimals have not been
fully clarified.

\cite{Guillot+2014} proposed detaile expressions for the accretion
rates of dust grains in a protoplanetary gaseous disk for a wide range
of dust and planetesimal sizes.  They point out that, when using the
numerical results by \cite{SekiyaTakeda2003}, the accretion
probability drops off by several orders of magnitude at relatively
small dust grains (within what they refer to as the ``hydrodynamical
regime'').  This strong depletion can be easily explained by
qualitative physical arguments.  The motion of small grains is
strongly coupled to the gas flow. \cwosug{They follow the streamlines
  around the large object and are hence deviated from possible}{Since
  they follow gas streamlines they may avoid} (head-on) collisions
with the planetesimal.  However, the numerical simulations by
\cite{SekiyaTakeda2003} assume laminar flow, while it is known that
the disk gas is most likely to be turbulent.  Observationally inferred
accretion rate in T-Tauri disks is far higher than that caused by
molecular viscosity.  Consequently, protoplanetary disks are expected
to be turbulent, with a turbulent eddy viscosity $\nut$ much higher
than the molecular kinematic viscosity.  Turbulence should affect the
reduction in the dust accretion probability in hydrodynamical regime.
Recently, \cite{mitra-wettlaufer-etal:2013} studied the influence of
turbulence on the dust impact velocity by means of two-dimensional
direct numerical simulations. They found a velocity distribution with
exponential tails and argued that most of the dust collides with a
speed comparable to that of the head wind.

The purpose of this paper is to propose expressions for dust accretion
probabilities \cwoadd{on planetesimals} in turbulent gas, based on three-dimensional direct
hydrodynamical simulations. Gravity effects are omitted in this study
and are left for a future work. We follow the approach introduced
by~\cite{homann2015concentrations} and consider an idealized situation
in which the planetesimal is assumed to be spherical with a smooth
surface. The gas is modeled as a purely hydrodynamical and
incompressible fluid so that other possible modes originating from
either the magneto-rotational instability (MRI) or compressibility are
not taken into account. Additionally, we neglect any shear, disregard
gravitational effects, and do not allow for rotation of the
planetesimal. The aim of this work is twofold: First, it shall provide
a detailed study of collisions between small inertial grains and a
large spherical object and, secondly, it shall determine its
consequences on the accretion of dust by planetesimals in the context
of planet formation.

It is worth mentioning that our results, involving the hydrodynamical
interactions and the collisions between a large spherical inclusion
and small particles with inertia, have important applications in
contexts that go beyond planet formation.  In atmospheric physics,
this problem is known as ``inertial impaction'' and is relevant for
estimating rates in rain formation or wet deposition of aerosols where
a falling water drop scavenges smaller cloud droplets or solid
pollutants. Studies of such problems often rely on collision
efficiencies and, again, mainly formulas from laminar flow conditions
are used \citep{berthet-etal:2010}. A popular formula is that
of~\cite{slinn:1974}, who proposed a fit of the collision efficiency
for the inertial impaction regime. Later, he included molecular
diffusion and extended his formula to smaller
projectiles~\citep{slinn:1976,slinn:1983}.  Inertial impaction is also
important for the design and improvement of industrial
filters. \cite{haugen-kragset:2010} studied the impact of particles on
a cylinder in a two-dimensional laminar inflow. Later,
\cite{rivedal2011} investigated the case of a turbulent inflow and
found that turbulent fluctuations yield up to ten times more
collisions than a corresponding laminar inflow.

The paper is organized as follows.  In
Section~\ref{sec:protoplanetary_turbulence}, we explain the disk model
and notations.  In Section~\ref{sec:numerical_model}, we describe the
numerical method that we use in our approach.  In
Section~\ref{sec:results}, we present the results of our
hydrodynamical simulations. In particular, we propose an expression
for the accretion probability in laminar and turbulent flows as a
function of both the Stokes number, which compares dust inertia to the
perturbation of the gas motion due to the planetesimal, and the
turbulent intensity of the surrounding gas flow.
Section~\ref{sec:astro} is devoted to astrophysical discussions on the
obtained results.  Section~\ref{sec:conclusions} contains a summary
and some concluding remarks.

\section{Protoplanetary disk turbulence model \& notations}
\label{sec:protoplanetary_turbulence}

\subsection{Disk model}
\label{sec:disk_model}
We consider a minimum mass solar nebula (MMSN) model
\citep{Weidenschilling1977i,Hayashi1981}, where the surface density
and temperature of the gas in the disk are given in terms of
power laws:
\begin{align}
  \Sigma_\mathrm{gas} = \Sigma_1 \left( \frac{r}{\mathrm{au}} \right)^{-3/2}, \\
  T_\mathrm{gas} = T_1 \left( \frac{r}{\mathrm{au}} \right)^{-1/2}.
\end{align}
$\Sigma_1$ and $T_1$ are the values at 1 a.u., for which we choose
$\Sigma_1=1700\ \mathrm{g\ cm^{-2}}$ and $T_1=270$ K. \MODHOL{(\cwosug{We note here that a}{A} list of symbols and their definitions used in this paper
  is summarized in the appendix \ref{appendix}.)} Assuming the disk is
vertically isothermal, with the isothermal sound speed $c_s=\sqrt{k_B
  T_\mathrm{gas}/\mu} \propto r^{-1/4}$ and a mean molecular weight of
$\mu=2.34 m_H$, the density reads
\begin{equation}
    \rho(r,z) = \frac{\Sigma_\mathrm{gas}(r)}{H\sqrt{2\pi}} \exp\left[
      -\frac{1}{2} \left( \frac{z}{H} \right)^2 \right],
    \label{eq:rho-rz}
\end{equation}
where $H=c_s/\Omega_\mathrm{K}$ is the disk scale height and
$\Omega_\mathrm{K}\equiv\sqrt{GM_\star/r^3}$ the orbital (Keplerian) frequency at
distance $r$ from the star. It follows that $H/r\propto r^{1/4}$, so
that the disk is flared.

Due to the density and temperature gradients, the disk is slightly
supported by pressure. The amount of pressure support $\Delta g =
(\mathrm{d}P/\mathrm{d}r)/\rho$ compared to the solar gravity is often
expressed in terms of a parameter $\eta$ defined as
\begin{eqnarray}
  \eta \equiv -\frac{\Delta g}{2g_\star} 
  &=& -\frac{P}{2\rho\, \Omega_\mathrm{K}^2\, r^2} \frac{\mathrm{d}\log P}{\mathrm{d}\log r}
      = \frac{1}{2}\,\left( \frac{c_s}{\v_\mathrm{K}} \right)^2 \nabla_{\rm log} P\nonumber \\
  &=& \frac{1}{2}\,h^2\,\nabla_{\rm log} P
      \approx 1.63\, h^2,
\end{eqnarray}
where we introduced the pressure logarithmic gradient $\nabla_{\rm
  log} P \equiv -\mathrm{d}(\log P)/\mathrm{d} (\log r)$, the
Keplerian velocity $\v_\mathrm{K}\equiv\Omega_\mathrm{K}\, r$, and the
disk aspect ratio $h\equiv c_s/\v_\mathrm{K}=H/r$. The numbers are
obtained from an MMSN disk with the above power law profiles for
density and temperature. It follows that the motion of the disk is
less than Keplerian \citep{Weidenschilling1977}; the gas flows at a
speed equal to $(1-\eta)\,\v_\mathrm{K}$. Also, the headwind that is
faced by a big body (a.k.a. planetesimal), which moves at a Keplerian
speed through the sub-Keplerian rotating gas, then reads
\begin{equation}
    U_c = \eta\,\v_\mathrm{K} = 52\ \mathrm{m\ s^{-1}}
\end{equation}
which, for an MMSN disk is independent of the distance $r$ to the
star.

\subsection{Turbulence model}
\label{sec:turbulence_model}
The gas has a mean-free path of
$\ell_\mathrm{mfp}=\mu/(\sqrt{2}\,\sigma_\mathrm{mol}\,\rho)$ where
$\sigma_\mathrm{mol}\approx 2\times10^{-15}\ \mathrm{cm^2}$
\citep{ChapmanCowling1970} is the molecular cross section. The
kinematic molecular viscosity of the gas is
$\nu_\mathrm{mol}=(1/2)\,\v_\mathrm{th}\,\ell_\mathrm{mfp}$ where
$\v_\mathrm{th}=\sqrt{8/\pi}\,c_s$ is the mean molecular thermal
speed.

However, for these parameters the resulting diffusion timescales
$\sim r^2/\nu_\mathrm{mol}$ are too long to explain, for example, the
measured accretion rate in T-Tauri disks. Consequently, protoplanetary
disks are expected to be mildly turbulent, but still subsonic, with a
turbulent eddy viscosity $\nut$ assumed to be much larger
than the molecular kinematic viscosity. An often used parameterization
is the alpha-viscosity model of \citet{ShakuraSunyaev1973}:
\begin{equation}
  \nut \approx \alpha\,c_s\,H.
    \label{eq:ss-alpha}
\end{equation}
The dimensionless constant $\alpha$ is a disk-dependent parameter. Its
value may range from a minimum of $\alpha\approx 10^{-5}$, when the
turbulence originates from the Kelvin-Helmholtz instability of a dust
layer at the mid-plane, to perhaps $\alpha\approx 0.1$ for the most
violent disks with a strong magneto-rotational instability (the MRI;
\citealt{BalbusHawley1991}). Generally, the \cwosug{strength of the MRI}{saturation level of the MRI turbulence}
depends on the magnitude of the magnetic field that threads the disks,
the level of ionization, and the amount of dust \cwoadd{\citep{Turner+2014}}.
Consequently, a
variety of $\alpha$-values can be expected
\citep{OrmelOkuzumi2013}. But even for a weak level of turbulence \cwoadd{-- meaning, $\alpha\ll1$ --} the
Reynolds number of the flow, $Re=\nut/\nu_\mathrm{mol}$, is
\cwoadd{very} large \cwoadd{by virtue of the very low densities that characterize astrophysical environments}. \cwosug{We indeed have for an MMSN disk}{Indeed, for an MMSN density profile:}
\begin{equation}
  Re = 2\,\alpha\,\frac{H}{\ell_\mathrm{mfp}} \approx
  2.4\times10^{12}\,\alpha\,\left( \frac{r}{1\ \mathrm{au}}
  \right)^{-3/2}.
\end{equation}

For such expected large values of the Reynolds number, one can
reasonably assume that the gas flow is in a developed
three-dimensional turbulent regime which can be described using
\citet{Kolmogorov1941} phenomenology. The kinetic energy is injected
at a large (integral) scale $L$ by a hydro- or magnetohydrodynamical
instability. In both cases the injection mechanism is associated to
the sub-Keplerian shear. The associated shear rate sets the typical
turbulent large-eddy turnover time to
$t_L \propto \Omega_\mathrm{K}^{-1}$ \citep{CuzziEtal2001}. The eddy
viscosity then reads
$\nut = L^2/t_L = L^2\,\Omega_\mathrm{K}$ leading, together
with Equation (\ref{eq:ss-alpha}), to $L=\alpha^{1/2}\,H$ and
$\v_L \equiv L/t_L = \alpha^{1/2}\,c_s$. Note that the assumptions of
incompressibility ($\v_L \ll c_s$) and three-dimensionality ($L\ll H$)
of the turbulent flow in the disk are both fulfilled as long as
\cwosug{$\alpha^{1/2} \ll 1$}{$\alpha \ll 1$}. According to \citet{Kolmogorov1941}
phenomenology, the kinetic energy cascades downscale with a constant
rate
$\varepsilon_\mathrm{Kol} = \v_L^2/t_L =
\alpha\,c_s^2\,\Omega_\mathrm{K}$
until it reaches the smallest active scales, the Kolmogorov scale
$\lkol$ below which it is dissipated by molecular viscosity.  The
typical velocity $\v_\ell$ of eddies whose size $\ell$ lies in the
inertial range $\lkol\ll\ell\ll L$ reads
$\v_\ell\simeq (\varepsilon_\mathrm{Kol}\,\ell)^{1/3}$. The Kolmogorov
length is defined as the scale where the scale-dependent Reynolds
number $Re(\ell) \equiv \v_\ell\,\ell / \nu_\mathrm{mol}$ is unity, so
that $\lkol=\nu_\mathrm{mol}^{3/4}/\varepsilon_\mathrm{Kol}^{1/4}$ and
the associated timescale
$t_\mathrm{Kol} =
\nu_\mathrm{mol}^{1/2}/\varepsilon_\mathrm{Kol}^{1/2}$.
The integral Reynolds number $Re = \nut/\nu_\mathrm{mol}$
gives the extension of the inertial range: $L/\lkol = Re^{3/4}$ and
$t_L/t_\mathrm{Kol} = Re^{1/2}$.  

\subsection{Dimensionless quantities}
Let us now turn back to our original problem, that is the
aerodynamic interactions between the disk gas flow and a solid
planetesimal of size $R_{\rm p}$, diameter $d=2R_{\rm p}$.  There are three dimensionless quantities characterizing
the system: (the numerical values below are those obtained for an MMSN disk)\\
\begin{itemize}
    \item The \emph{planetesimal Reynolds number} 
    \begin{equation}
      \Rep \equiv \frac{dU_c}{\nu_\mathrm{mol}}
      = \frac{\nabla_\mathrm{log}P\,h}{\sqrt{8/\pi}} \left( \frac{d}{\ell_\mathrm{mfp}} \right)
      \approx 2.7\times10^4 \left( \frac{d}{\mathrm{km}} \right) \left(
        \frac{r}{1\ \mathrm{au}} \right)^{-2.5},
      \label{eq:Rep}
    \end{equation}
    which corresponds to the strength of inertia with respect
    to molecular viscous forces for the gas flow surrounding the
    planetesimal and characterizes how turbulent is its wake.\\

    \item The \emph{turbulent intensity}
    \begin{equation}
      I \equiv \frac{\v_L}{U_c}
      = \frac{\alpha^{1/2}\,c_s}{(1/2)\,h^2\nabla_P \v_K}
      =  \frac{\sqrt{4}\alpha^{1/2}}{h\,\nabla_\mathrm{log}P}
      \approx 20\, \alpha^{1/2} \left( \frac{r}{1\ \mathrm{au}} \right)^{-1/4},
    \end{equation}
    which measures how strong the turbulent velocity fluctuations are
    compared to the speed of the planetesimal slip.\\ 
    
    \item the ratio
    between the planetesimal size and the Kolmogorov scale
    \begin{equation}
      \begin{split}
        \delta_{_\mathrm{Kol}}\equiv\frac{d}{\lkol} & = \frac{d}{\alpha^{1/2}H}
        Re^{3/4} = 2^{3/4}\alpha^{1/4} \left( \frac{d}{H} \right) \left(
        \frac{H}{\ell_\mathrm{mfp}} \right)^{3/4} \\ & \approx
        400\,\alpha^{1/4} \left( \frac{d}{\mathrm{km}} \right) \left(
        \frac{r}{1\ \mathrm{au}} \right)^{-19/8}
      \end{split}
    \end{equation}
    which \JB{measures the range of turbulent eddies that might interfere
      with the planetesimal perturbation of the gas flow. It is indeed
      known that the flow perturbation occurs on scales of the order of
      $d$, so that all turbulent eddies of sizes between $\lkol$ and $d$
      can potentially modify the gas flow around the spherical
      planetesimal.}

\end{itemize}

\begin{figure}
\centering
\includegraphics[width=\hsize]{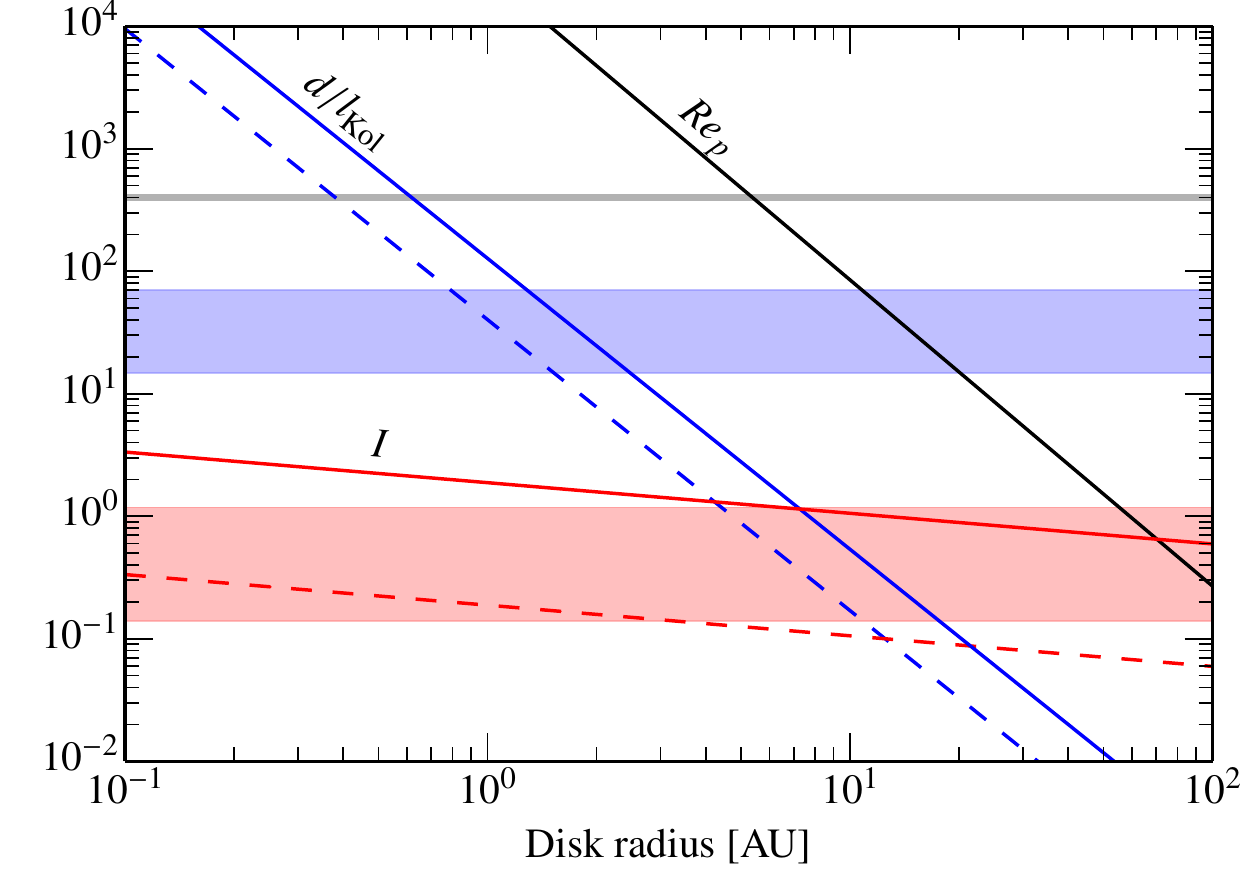}
\caption{\label{fig:mmsn} Turbulent disk parameters: Planetesimal
  Reynolds number $\Rep$ (black), dimensionless turbulent intensity
  $I$ (red), and planetesimal to Kolmogorov scale $d/\lkol$ (blue) for
  a MMSN disk model, disk turbulent parameters $\alpha=10^{-4}$
  (dashed) and $\alpha=10^{-2}$ (solid) and planetesimal diameter
  $d=1$\,km. The blue and black line thus scale up and down with the
  planetesimal size. The shaded regions are covered by the numerical
  experiments (see Table~\ref{table2}).}
\end{figure}

Figure~\ref{fig:mmsn} shows the dependence of the various parameters
with the orbital distance for two values of $\alpha$ compared to the
assumptions used for the numerical calculations. We can see that
$\Rep$ decreases with increasing orbital distance (kinematic viscosity
goes up). Similarly, $d/\lkol$ decreases because $\lkol$ increases
(turbulent scales are getting larger). The turbulent intensity is not
a very strong function of disk radius, but it depends quite strongly
on the turbulence parameter $\alpha$.

In the laminar case, $\Rep=400$ corresponds to a 1\,km planetesimal
located at $\sim 5$\,au in the MMSN disk, or e.g. to a 10\,km
planetesimal at $\sim 12\,$au.

For the turbulent case, ideally we should find parameters for which
all three lines match with parameters of the numerical
experiments. However, because of the $\sim 10^{6}$ mismatch between
the Reynolds number in the disk and the maximum one that can be
attained by present-day numerical simulations, this is not yet
possible. We come back to this issue when applying the results of
numerical models to real disk conditions.

\section{Numerical model}
\label{sec:numerical_model}

\subsection{Gas flow}
We focus on the collision rate of a stream of dust particles with one
spherical planetesimal. To study this situation we perform 3D direct
numerical simulations (DNS) of a hydrodynamic flow around a spherical
object with a no-slip boundary conditions at its surface. The overall
method consists in a combination of a standard pseudo-Fourier-spectral
solver with a penalty method that is explained in this section.

We integrate the incompressible Navier-Stokes equations
\begin{equation}
  \label{eq:navier-stokes}
  \partial_t {\bm u} = - \bm u\cdot\nabla\bm u -
  \frac{1}{\rho_\mathrm{g}}\nabla p +\nu_\mathrm{mol} \nabla^2 {\bm u} + {\bm f},
  \quad\nabla \cdot {\bm u} = 0,
\end{equation}
for the gas velocity ${\bm u}$, where $\rho_\mathrm{g}$ is the gas
density, $\nu_\mathrm{mol}$ its kinematic molecular viscosity and
${\bm f}$ a force. The latter maintains a uniform inflow speed and an
eventually ambient turbulent flow. Its form is described later. The
pseudo-Fourier-spectral approach consists in computing spatial
derivatives in Fourier-space and convolutions arising from the
non-linear terms in real space. A Fast-Fourier transform is used to
switch between the two spaces. We use the P3DFFT library (see
\citealt{p3dfft_paper:2012}) that is very efficient on massive
parallel computers.

The Navier--Stokes equation is integrated in the reference frame of
the planetesimal. The spatial average of the velocity field is thus
fixed and given by the planetesimal speed $U_c$ relative to the gas.
Equation (\ref{eq:navier-stokes}) is associated with a no-slip
boundary condition at the surface of the planetesimal, which is
assimilated to a spherical object at rest whose diameter is denoted by
$d$ and its center by $\bm X_S$. We thus have
\begin{equation}
  \label{eq:bc}
  {\bm u}(\bm x, t) = 0 \quad\mbox{for }\ |\bm x - \bm X_S| = {d}/2.
\end{equation}
Numerically, this no-slip condition is enforced by an immersed
boundary technique (IBM). \MODHOL{The \cwosug{later}{latter} technique was first used
  to simulate the blood in flow in the context of a human heart by
  \cite{peskin-1972}. Today, a variety of different approaches exists
  \citep{mittal-iaccarino:2005}. Generally speaking, IBM consists in
  solving fluid equations in domains with complex boundary conditions
  such as moving heart valves on Cartesian grids. As such boundaries
  are generally not grid conform, their effect on the flow has to be
  modeled. For our problem of a spherical obstacle at rest}, the idea
consists in defining the velocity field in the full domain enforcing a
vanishing velocity in the entire object, that is for all $\bm x$ such
that $|\bm x - \bm X_S| \le {d}/2$. The Navier-Stokes equation
(\ref{eq:navier-stokes}) then has to be modified by introducing in its
right-hand side a penalty force $\bm f_ b(\bm x, t)$, which acts as a
Lagrange multiplier associated to the constraint defined by the
boundary condition (\ref{eq:bc}). The full problem
(\ref{eq:navier-stokes})-(\ref{eq:bc}) can then be rewritten as
\begin{equation}
  \label{eq:ib}
  \partial_t {\bm u} = L(\bm u) +\bm f_b, \quad\nabla \cdot
  {\bm u} = 0,
\end{equation}
where $L(\bm u)$ denotes the right hand side of
(\ref{eq:navier-stokes}). In order to compute $\bm f_b$ we make use of
a direct forcing method introduced by \cite{fadlun-verzicco-etal-2000}
where we directly impose the planetesimal velocity to the grid using
the technique of a pressure predictor and an improved modeling of the
spherical inclusion. Benchmarks (see ~\citealt{homann-bec-etal:2013})
of this method for a fixed sphere show good agreement with existing
data. This method has also been used for the study of moving neutrally
buoyant particles in homogeneous isotropic turbulence in
\cite{homann-bec:2010} and \cite{cisse-homann-etal-2013}. \MODHOL{A
  similar IBM has been used by \cite{uhlmann:2005} together with
  second order finite-difference Navier-Stokes solver to simulate
  turbulent suspension involving many particles.}

\subsection{Dust particles}
Dust particles are modeled by spherical inertial particles with a
radius $a$ much smaller than the smallest scales of the flow, so that
they can be approximated by point particles. Further, we assume that
these particles move sufficiently slow with respect to the gas and
that their mass density $\rho_\mathrm{p}$ is much higher than the gas
density $\rho_\mathrm{g}$. With these assumptions the dominant
hydrodynamic force exerted by the gas is a drag force, which is
proportional to the velocity difference between the particle and the
gas flow (see \citealt{maxey-riley:1983,gatignol:1983})
\begin{equation}
  \ddot{\bm X} = \frac{1}{\taus}\! \left[\bm u(\bm X\!,t) \!-\!
  \dot{\bm X} \right] 
  \label{eq:particles}
\end{equation}
where the dots stand for time derivatives.  $\taus$ is called the
response (or stopping) time and is a measure of particle inertia. It
is the relaxation time of the particle velocity to that of the gas
\citep[see][for a description of how the particle size relates to the
  stopping time]{Guillot+2014}.
Finally, we also assume that the particles are sufficiently diluted to
neglect any interaction among them and any back-reaction on the flow.

Usually, particle inertia is quantified in terms of the Stokes number
$\St = \taus\, / \, \tauc$ defined by non-dimensionalizing their
response time by a characteristic time scale $\tauc$ of the carrier
flow. The present problem involves different relevant time scales.
$T_d=d/U_c$, the time it takes a dust particle to pass the
planetesimal, $t_L$, the turbulent large-eddy turn-over time and
$t_\mathrm{Kol}$, the turbulent dissipation time scale. If not
otherwise specified, we use $\tauc=T_d$ as this time scale rules the
collision efficiency in the laminar regime: Particles with small $\St$
are closely coupled to the flow and are swept around the
obstacle. Large $\St$ particles preferentially collide with
it.\footnote{Note that in the context of astrophysical disks, a
  different Stokes number is often defined from $\tau=\taus\Omega_{\rm
    K}$.  The definition of $\St$ that is used here is denoted by
  $\tau_{\rm f}$ in e.g., \cite{SekiyaTakeda2003} and
  \cite{Guillot+2014}.}

In studies concerned with the small-scale dynamics of inertial
particles one usually uses
$\tauc=t_\mathrm{Kol}$~\citep{bec-etal:2006} and we denote the
associated Stokes number $\St_\mathrm{Kol}$.

\subsection{Simulation setup}

\begin{figure*}
\sidecaption
\includegraphics[width=0.7\hsize]{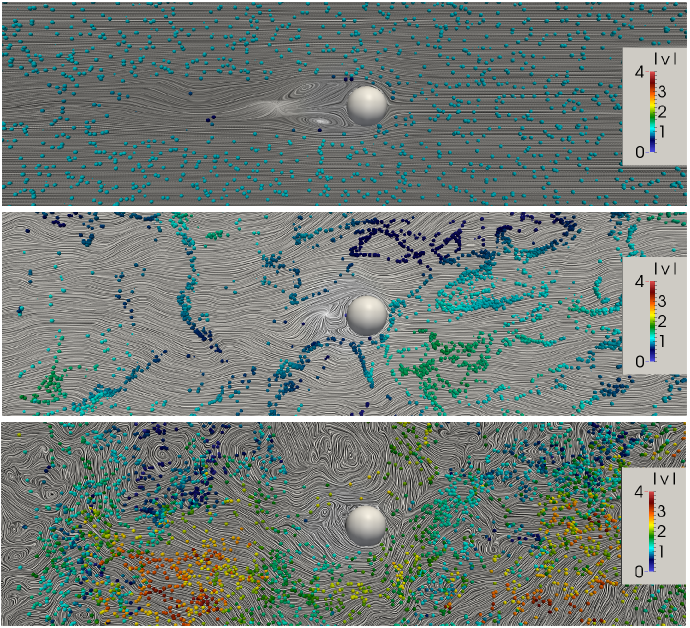}
\caption{\label{fig:flow_particles}Instantaneous streamlines of the
  gas flow (entering from the right with a speed $U_c$) around a
  planetesimal with $Re_p=400$ and dust particles ($\St=0.8$) for
  $I=0$ (laminar, $\delta_\mathrm{Kol}=0$), $I=0.29$ (moderately
  turbulent, $\delta_\mathrm{Kol}=25$), and $I=1.18$ (strongly
  turbulent, , $\delta_\mathrm{Kol}=70$) from top to bottom. }
\end{figure*}

We performed simulations of a planetesimal moving through a uniform
and different turbulent flows.  The physical situation is illustrated
in Fig.~\ref{fig:flow_particles} where the planetesimal, together with
dust particles are shown in a small slice. These are snapshots taken
from our three-dimensional simulations. The planetesimal experiences a
headwind speed $U_c$ from the right that is advecting the dust
particles. The upper panel is taken from a simulation with a uniform
inflow while the two others include turbulence with two different
intensities. Dust particles that have collided with the planetesimal
are missing downstream, resulting in an empty region behind the
planetesimal that is clearly visible for the laminar flow, but much
less in the turbulent cases.

These figures show important differences between laminar and turbulent
gas flows. The spatial distribution of the dust as well as its local
velocity strongly depend on the carrier flow type. Note that while
$\St=0.8$ for all cases shown in Fig.~\ref{fig:flow_particles},
$\St_\mathrm{Kol}=1.25$ for $I=0.29$ and $\St_\mathrm{Kol}=9.8$ for
$I=1.18$ so that the clustering properties of the particles change
from one value of $I$ to the other.

The runs involving a turbulent flow are set up in the following way
(see table~\ref{table1} for parameters and definitions). An initially
smooth large scale flow is integrated according to
(\ref{eq:navier-stokes}) without any forcing $\bm{f}$.  Once a
turbulent flow has developed (after approx. $1\!-\!2\,t_L$) the
velocity field is forced by keeping constant the energy content of the
two lowest wave number shells ($1\le k \le 2)$ in Fourier space. This
leads to a statistically stationary turbulent flow to which the
planetesimal is added. For this, the root-mean-square value $\v_{L}$
of the velocity fluctuations is normalized to the values given in
table \ref{table1} for each simulation. A mean velocity of $U_c$ in
one direction is imposed by keeping constant the zero Fourier mode of
the corresponding component of the velocity. The planetesimal is
modeled via the mentioned immersed boundary method. The integration is
continued until a statistically stationary state is reached again. At
this point, inertial particles are seeded at a constant rate into the
flow in a plane sufficiently far from the planetesimal so that they
have enough time ($> \taus$) to relax to the flow. During the
simulation we remove and record all the particles that are touching
the spherical planetesimal or reaching the end of the computational
domain. On average the domain is filled with approximately ten million
particles.

The laminar flow simulations (see table~\ref{table2} for parameters
and definitions) start with a uniform flow ($U_c=1$) in which the
planetesimal is placed. Disturbances produced by its wake are removed
at the end of the computational domain via another application of the
penalty method, so that they are not re-injected upstream the
planetesimal by the periodic boundary conditions. The dust particles
are introduced into the flow once a (statistically) stationary state
is reached.

The time integration of (\ref{eq:navier-stokes}) uses a Runge-Kutta
scheme of third order. The grid resolution is chosen in order to
resolve all small scales of the problem: those of the spherical
planetesimal boundary layer, all turbulent scales in its wake and the
smallest scales of the possibly turbulent ambient flow.

All physical flow parameters are determined by three dimensionless
parameters: the Reynolds number of the planetesimal
$Re_p=U_c\,d/\nu_\mathrm{mol}$, the Reynolds number of the gas
$Re=\v_{L}\,L/\nu_\mathrm{mol}$, $L$ being the integral scale of the
ambient turbulent flow, and the turbulent large-scale intensity
$I=\v_{L}/U_c$. The latter measuring the strength of the large-scale
turbulent fluctuations compared to the mean flow velocity. In the
laminar case ($Re=0$, $I=0$), we varied the planetesimal Reynolds
number $Re_p$ from 100 to 1000. We analyze the effect of turbulent
fluctuations for one specific choice of the planetesimal Reynolds
number, namely $Re_p=400$ (turbulent wake) and vary $I$ from 0.14 to
1.18. The corresponding flow Reynolds numbers $Re$ (listed in
Table~\ref{table1}) are a consequence of our particular choice of the
external force. Freezing the energy content of the lowest shells in
spectral space does not allow for changing $L$ but only $\v_{L}$ that
enters in the definitions of both $I$ and $Re$.

The particle dynamics is characterized by the Stokes number $\St$. In
all simulations we consider streams of heavy dust particles with
response times $\taus$ in between 0.04 and 81.92, corresponding to
Stokes numbers $\St$ in the range $0.05\le \St\le63$. The main
parameters of all simulations are summarized in table~\ref{table1} and
table~\ref{table2}.

\begin{table*}
  \caption{\label{table2} Parameters of the turbulence simulations.
    $Re_p=U_c\,d/\nu_\mathrm{mol}$: planetesimal Reynolds number, $Re
    = \v_{L}L/ \nu_\mathrm{mol}$: outer gas flow Reynolds number,
    $\v_{L}$: root-mean-square velocity, $\varepsilon_\mathrm{Kol}$:
    mean kinetic energy dissipation rate, $\nu_\mathrm{mol}$:
    kinematic viscosity, $d$: planetesimal diameter, $\lkol
    =(\nu_\mathrm{mol}^3/\varepsilon_\mathrm{Kol})^{1/4}$: Kolmogorov
    dissipation length scale, $t_\mathrm{Kol} =
    (\nu_\mathrm{mol}/\varepsilon_\mathrm{Kol})^{1/2}$: Kolmogorov
    time scale, $L = \v_L^{3}/\varepsilon_\mathrm{Kol}$: integral
    scale, $t_L = L/\v_L$: large-eddy turnover time, $N^3$: number of
    collocation points.}  \centering
 \begin{tabular}{ccccccccccccccc}
   \hline\hline
   $Re_p$ & $Re$ & $U_c$ &$\v_L$&$\varepsilon_\mathrm{Kol}$&$\nu_\mathrm{mol}$ & d& $\lkol$&$t_\mathrm{Kol}$&$L$   &$t_L$& $N_x\times N_y\times N_z$ & $N_p$\\
   \hline 
   400    & 200 & 1  & 0.14          &$9.2\cdot 10^{-4}$  &0.002& 0.8 &$0.054$ & $1.46$   &2.92 & 21.0 &$256 \times 256 \times 2048$&$\approx 10^7$ \\
   400    & 450 & 1  & 0.29          &$7.7\cdot 10^{-3}$  &0.002& 0.8 &$0.032$ & $0.51$   &3.04 & 10.6&$256 \times 256 \times 2048$&$\approx 10^7$ \\
   400    & 600 & 1  & 0.39          &$1.95\cdot 10^{-2}$ &0.002& 0.8 &$0.025$ & $0.32$   &3.09 & 7.9 &$256 \times 256 \times 2048$&$\approx 10^7$ \\
   400    & 900 & 1  & 0.60          &$6.9\cdot 10^{-2}$  &0.002& 0.8 &$0.018$ & $0.17$   &3.10 & 5.2 &$256 \times 256 \times 2048$&$\approx 10^7$ \\
   400    & 3000 & 1  & 1.18         &$4.8\cdot 10^{-1}$  &0.002& 0.8 &$0.0114$ & $0.065$ &3.40 & 2.9 &$256 \times 256 \times 2048$&$\approx 10^7$ \\
   \hline 
 \end{tabular}
\end{table*}

\begin{table}
  \caption{\label{table1}Parameters of the laminar simulations.
    $Re_p=U_c\,d/\nu_\mathrm{mol}$: planetesimal Reynolds number ,
    $d$: planetesimal diameter, $\nu_\mathrm{mol}$: kinematic
    viscosity, $N_x\times
    N_y\times N_z$: number of collocation points; The following
    parameters apply to all the three simulations: lengths of the
    computational domain $l_x\times l_y\times l_z =
    2\pi\times2\pi\times16\pi$, inflow speed $U_c=1$, number of small
    particles $N_p\approx 10^7$} \centering
  \begin{tabular}{ccccccccrr}
    \hline\hline
    $Re_p$ & $d$ & $\nu_\mathrm{mol} $ & $N_x\times N_y\times N_z$ \\ \hline
    100  & 0.8  & $2\cdot 10^{-3}$ & $256 \times 256 \times 2048$  \\
    400  & 0.8  & $2\cdot 10^{-3}$ & $256 \times 256 \times 2048$  \\
    1000  & 0.65 & $6.5\cdot 10^{-4}$ & $512 \times 512 \times 4096$\\
    \hline 
  \end{tabular}
\end{table}

\section{Results}
\label{sec:results}

\subsection{Laminar settings}

\begin{figure}[h]
\centering
\includegraphics[width=0.8\hsize]{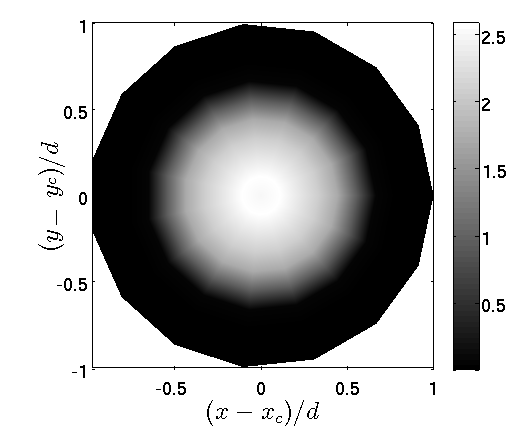}
\caption{\label{fig:accretionPosition} Projected frontal view of the
  average probability density of impaction on the planetesimal
  \MODHOL{for dust particles with $\St=0.2$}. White: Many collisions;
  Black: No collisions.}
\end{figure}

In a uniform gas flow all dust particles have the same velocity far
from the planetesimal. This physical situation is fully determined by
the Reynolds number of the planetesimal $Re_p$ and the Stokes number
$\St$ of the dust particles. A typical flow pattern is illustrated in
the upper panel of Fig.~\ref{fig:flow_particles} where a planetesimal
flies through a stream of dust particles while creating a moderately
turbulent wake. If an approaching dust particle collides with the
planetesimal or not is only determined by its Stokes number and impact
parameter. Without the hydrodynamic flow that deflects dust around the
obstacle it would just be the impact parameter ruling the collision
rate so that the hydrodynamic forces reduce the collision rate below
that of the geometric cross section.

\begin{figure}[h]
\centering
\includegraphics[width=0.9\hsize]{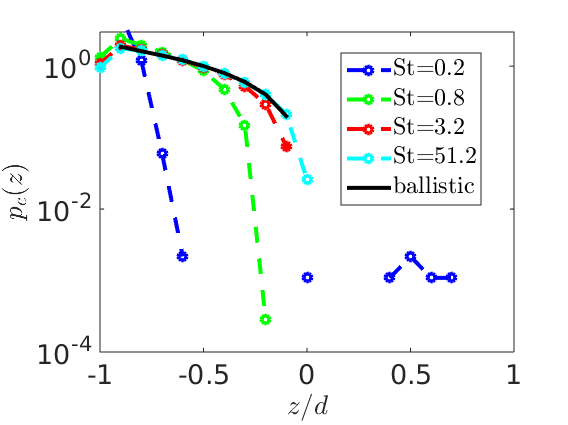}
\caption{\label{fig:accretionPosition_r}Probability density function
  of the stream-wise position $z$ at which colliding particles impact
  the planetesimal. The center of the planetesimal is located at $z=0$
  and separates its front (negative $z$) from its back (positive
  $z$). The 'ballistic' curve refers to straight dust trajectories
  that do not feel any hydrodynamic forces.}
\end{figure}

The colliding dust particles preferentially hit the planetesimal on
the axis of symmetry with a decreasing probability to its edge. Small
inertial particles only touch the planetesimal in a central region
leaving eventually an outer non-collisional ring (see
Fig.~\ref{fig:accretionPosition}). But this region already disappears
for Stokes numbers around unity, so that dust collisions fill the
complete front of the spherical planetesimal. \MODHOL{In the limit of
  infinite $St$ the distribution become uniform.} Rear collisions
virtually never happen (see Fig.~\ref{fig:accretionPosition_r}) in a
laminar flow that is because inertia prevents those particles that
moved around the planetesimal from getting into the recirculation
region of the wake. The few records observed might be numerical noise.

\begin{figure}[h]
\centering
\includegraphics[width=0.9\hsize]{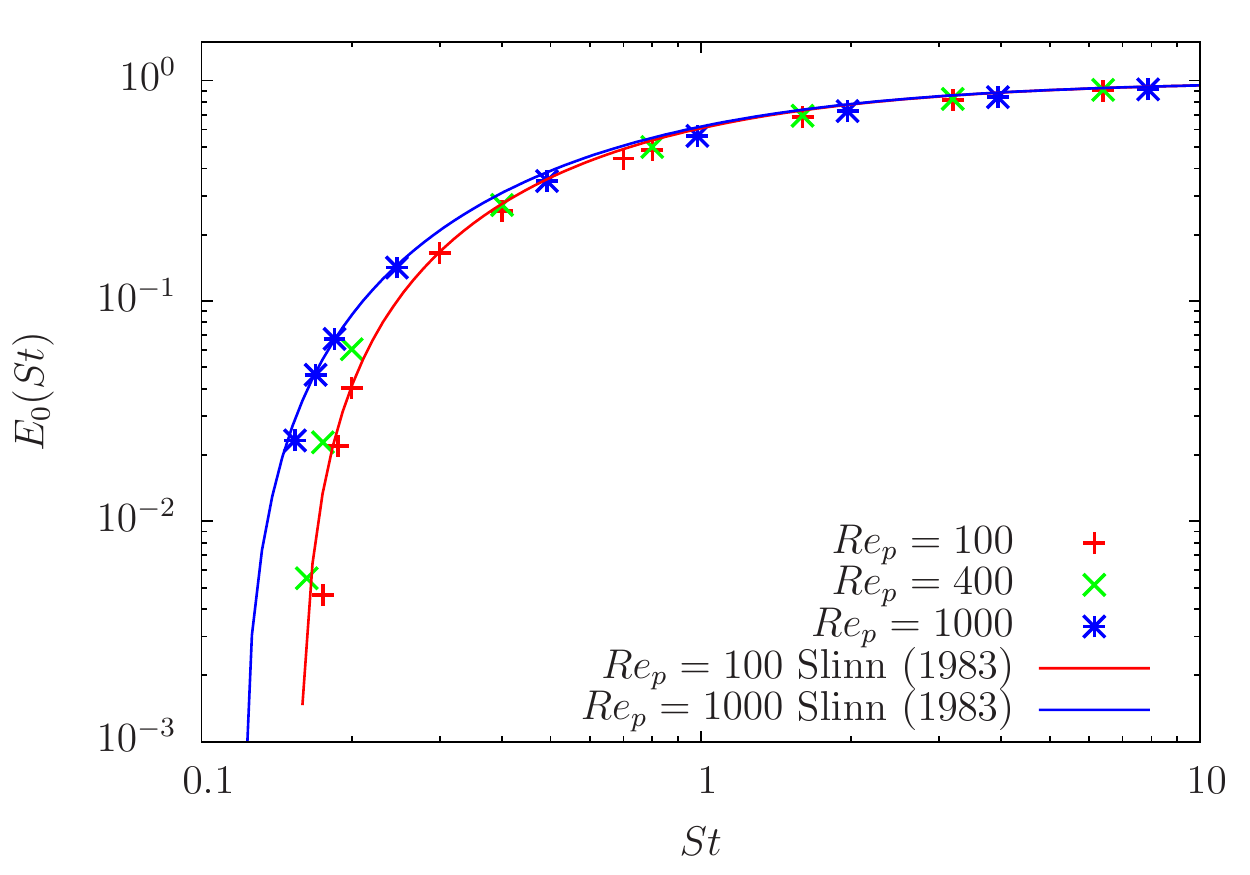}
\caption{\label{fig:collisionRate_comp} Collision efficiency as a
  function of the dust particle Stokes number obtained from direct
  numerical simulations with uniform inflows (symbols). The lines
  correspond to the fitting formula (\ref{eq:slinn}) proposed by
  \cite{slinn:1983}.}
\end{figure}

A central quantity of the accretion problem is the collision
efficiency (or collision rate) $E(\St)$ that is the ratio between the
actual collision rate and that expected for free-streaming particles
that do not feel any hydrodynamic force from the gas flow. The
\cwosug{later}{latter} rate is obtained from the geometrical
cross-section of the planetesimal and reads $n_{\rm
  d}\,U_c\,\pi\,d^2/4$, where $n_{\rm d}$ designates the number
density of dust particles. \MODHOL{We use the following notation:
  $E_0$ denotes the uniform gas flow case that we are going to analyze
  in this section and $E_I$ denotes the turbulent case parameterized
  by the turbulent intensity $I$ that is discussed in the following
  section.} The efficiency $E_0$ is a monotonously increasing function
of the dust particle Stokes number $\St$ that asymptotically reaches
unity for particles with very large inertia (see
Fig.~\ref{fig:collisionRate_comp}). The higher is $Re_p$ the more
probable are collisions, especially at small $\St$. In this low
inertia limit, $E_0$ sharply drops for all $Re_p$. The reason for this
is the well established existence of a critical Stokes number $\St_c$
below which no collisions occur at all \MODHOL{\citep{taylor:1940}}.

Indeed, in the case of an inviscid Euler flow (zero viscosity,
$Re_p=\infty$) it can be shown that $E_0$ drops to zero for small but
finite $\St^{\text{Euler}}_c=1/24$
\MODHOL{\citep{ingham-hildyard-etal:1990}}. The viscous boundary layer
of a finite $Re_p$ planetesimal that shrinks as $\sim 1/\sqrt{Re_p}$
reduces the collision probability and increases $\St_c$. For the
limiting case of a Stokes flow \cite{michael-norey:1970} computed
numerically a critical Stokes number of $0.605$.

\cite{slinn:1974,slinn:1976,slinn:1983} proposed a fitting function of
the collision rate of the form
\begin{equation}
\label{eq:slinn}
\begin{cases}
  E_0^{\text{Slinn}}(\St,\Rep)=\left(\frac{\displaystyle{\St-\St^{\text{Slinn}}_{\rm c}}}{\displaystyle{\St-\St^{\text{Slinn}}_{\rm c}+2/3}}
  \right)^{3/2} \\[1em]
  \St^{\text{Slinn}}_{\rm c}(\Rep)= \frac{\displaystyle{0.6+(1/24)\,
    \log(1+\Rep/2)}}{\displaystyle{1+\log(1+\Rep/2)}}.
\end{cases}
\end{equation}
$\St^{\text{Slinn}}_{\rm c}$ is thus a critical number below which the
(extremely low) collision probability is determined by different
physical mechanisms.  These expressions appear to have been fitted
empirically to both experimental and numerical data with uncertainties
of order $0.1$ on the determination of $E_0(\St)$
\cite[see][]{slinn:1974}.

\begin{figure}[h]
\centering
\includegraphics[width=0.9\hsize]{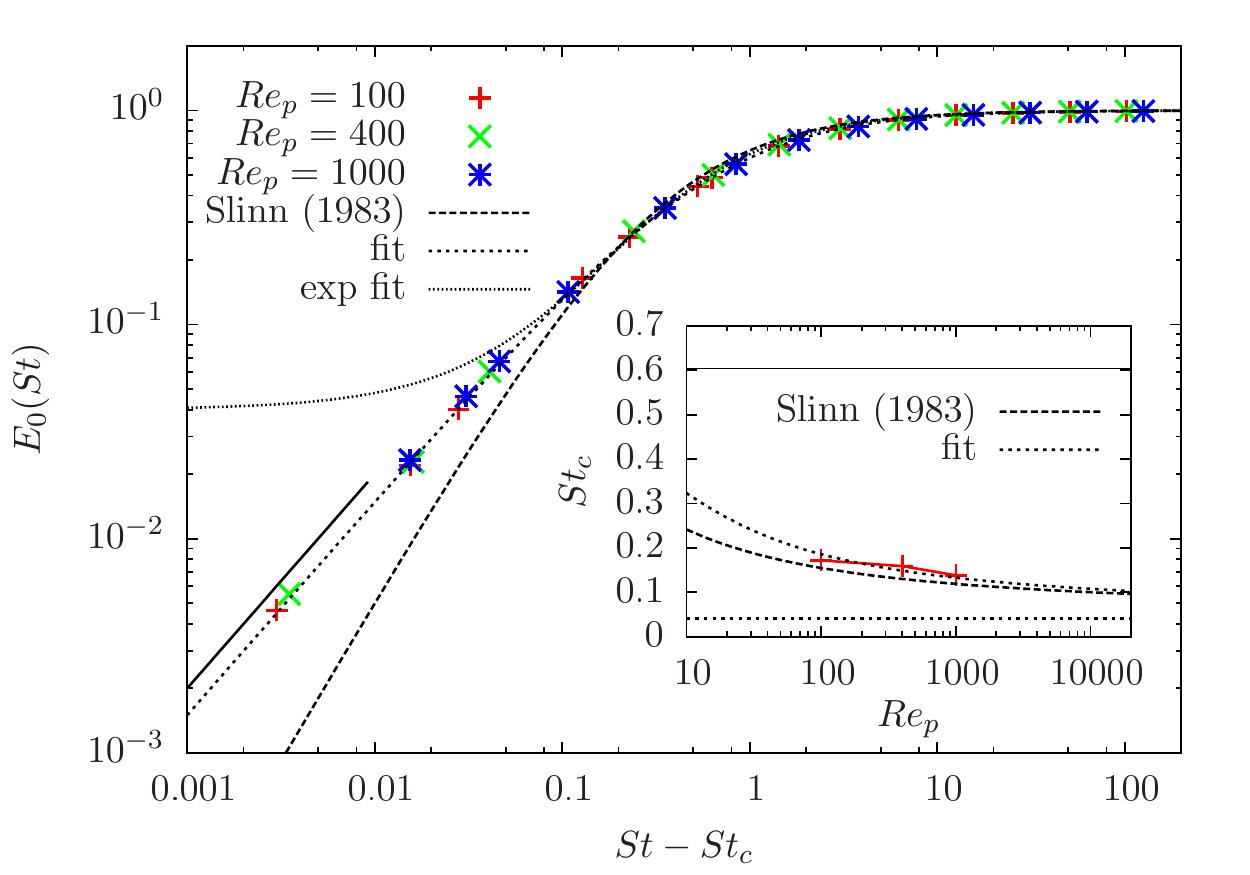}
\caption{\label{fig:collisionRate_Stc} Collision efficiency as a
  function of $\St-\St_c$, where $\St_c$ is the critical Stokes number
  below which no collision happen. fit=$(\St-\St_c)/(\St-\St_c+2/3)$
  (eq. (\ref{eq:e_fit})). \MODHOL{The solid line indicates a linear
    increase}. Inset: Critical Stokes number as a function of the
  planetesimal Reynolds number. fit=$\St^{\text{Slinn}}_c(\Rep/3)$.}
\end{figure}

It is useful to look in more detail at the small and large $\St$
efficiencies separately. In Fig.~\ref{fig:collisionRate_Stc} the
collision efficiency is shown as a function of $\St-\St_c$ and not
simply as a function of $\St$. We hence focus on the behavior of the
collision probability when approaching the critical Stokes
number. $\St_c$ is chosen in such a way that all curves fall on top of
each other so that differences for different Reynolds numbers
disappear. Our estimated $\St_c$ are close to that of Slinn
(\ref{eq:slinn}) (see inset of Fig.~\ref{fig:collisionRate_Stc}). The
case of large but finite values of $Re_p$ has also been addressed by
\cite{phillips-kaye:1999}, using matched asymptotics together with
numerical simulations of the particle dynamics inside the obstacle
boundary layer. They found critical Stokes numbers slightly larger
than those of Slinn (\ref{eq:slinn}), so that our value are in between
these two predictions.

$E_0$ increases linearly at small $\St-\St_c$ (see
Fig.~\ref{fig:collisionRate_Stc}), that is in contradiction to Slinn's
formula (\ref{eq:slinn}) that predicts $E_0\sim (\St-\St_c)^{3/2}$. To
incorporate this small-Stokes behavior we propose the fitting function
\begin{equation}
  \label{eq:e_fit}
\begin{cases}
  E_0(\St,\Rep)=\frac{\displaystyle{\St-\St_{\rm c}}}{\displaystyle{\St-\St_{\rm c}+2/3}} \\[1em]
  \St_{\rm c}(\Rep)= \frac{\displaystyle{0.6+(1/24)\,
      \log(1+\Rep/6)}}{\displaystyle{1+\log(1+\Rep/6)}}
\end{cases}
\end{equation} 
which is in excellent agreement with our data. Here we mention (as
already remarked by Slinn) that an exponential fit of the form
$\exp(-a/\St)$ also works quite well for not too small $\St$ but this
form does evidently not conform to a critical Stokes number. 

We also note that \citet{slinn:1974} had also proposed a similar fit
(linear instead of with a $3/2$ exponent), but opted for
eq.~\eqref{eq:slinn} which showed a marginal improvement to the
experimental and numerical data available at the time. The
uncertainties that we obtain here are much smaller and allow to
discriminate against the fit from eq.~\eqref{eq:slinn}.

\begin{figure}[h]
\centering
\includegraphics[width=0.9\hsize]{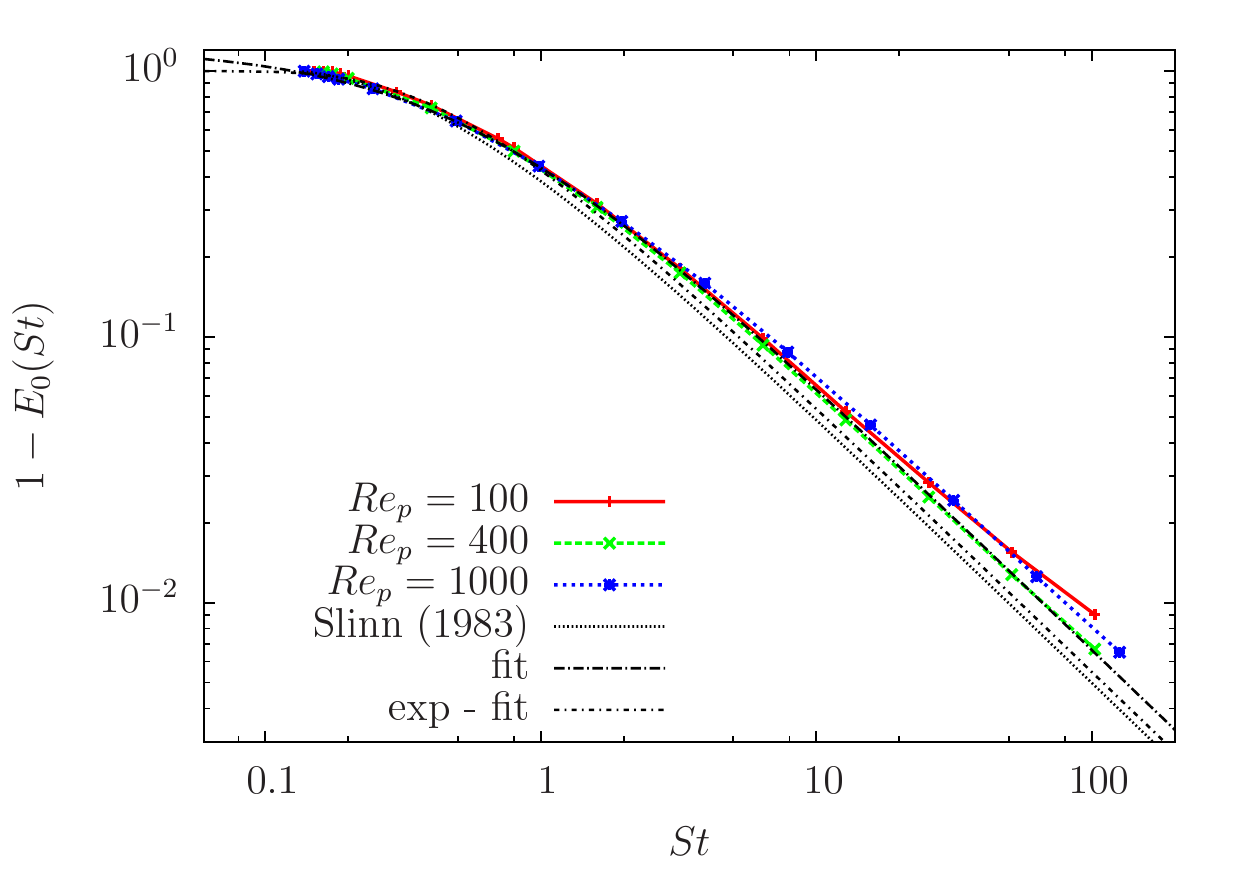}
\caption{\label{fig:collisionRate_largeSt} Collision efficiency
  deficit $1-E_0(\St)$ as a function of $\St$ for different planetesimal
  Reynolds numbers and the various fitting formula, as labeled.}
\end{figure}

To study the large-Stokes number behavior in details it is useful to
analyze $1-E_0(\St)$ that measures how the free-streaming particle limit
is recovered as a function of $\St$.  All curves for different $Re_p$
(see Fig.~\ref{fig:collisionRate_largeSt}) fall on the top of each
other and reveal a $\St^{-1}$ behavior at large $\St$. Again, we
observe slight deviations to Slinn's fitting formula, while our
proposed expression (\ref{eq:e_fit}) fairly matches the data.

\subsection{Turbulent settings}

One can expect that turbulent velocity fluctuations of the gas,
resulting in dust velocity fluctuations, alter the collision
statistics of dust and planetesimals. An analysis of these changes is
the subject of this section.

The turbulent accretion problem involves more parameters than the
laminar problem presented in the previous section. It is determined by
four dimensionless parameters. The planetesimal and dust properties
are of course still described by the Reynolds number $Re_p$ and the
dust Stokes number $\St$, but turbulence adds two additional
dimensionless parameters specifying the ambient turbulent flow. They
are the gas Reynolds number $Re=\v_L\,L/\nu_\mathrm{mol}$ and the turbulent
intensity $I=\v_L/U_c$ that compares the amplitude of turbulent
fluctuations with the mean velocity of the gas.

In this work, we explicitly vary $\St$ and the turbulent intensity $I$
and fix the planetesimal Reynolds number to $Re_p=400$ (weakly
turbulent wake) in order to limit computational costs. The gas
Reynolds number $Re$ is implicitly varied as it is coupled to $I$ due
to our specific forcing scheme that prescribes $L$ to approximately
half of the domain size.

The physical situation for two different turbulent intensities is
illustrated in the mid and bottom panel of
Fig.~\ref{fig:flow_particles}. Dust particles are in these cases
advected by a chaotic and irregular flow possessing coherent
structures, i.e. structures eventually persisting for a long
time. This has two important implications: First, the velocity of dust
is fluctuating spatially and temporally (see
Fig.~\ref{fig:flow_particles} b) and c)) and in turn (as we study in
details below) modifies the collision statistics. Second, from studies
of hydrodynamic turbulence it is known that inertial particles tend to
escape from coherent rotating regions of the flow and tend to cluster
in straining regions~\citep{squires-eaton:1991}. These agglomerations
are called preferential
concentrations\cite{shaw:2003,balachandar-eaton:2010}. However, we
expect that these concentrations play a negligible role for the
present study in which we are concerned with averaged accretion
quantities such as the collision efficiency. The reason is that the
temporally averaged dust concentration in the ambient flow is
approximately the same as in the laminar case (not shown) so that
particle density fluctuations average out.

\subsubsection{Collision efficiency}

\begin{figure}[h]
\centering
\includegraphics[width=0.9\hsize]{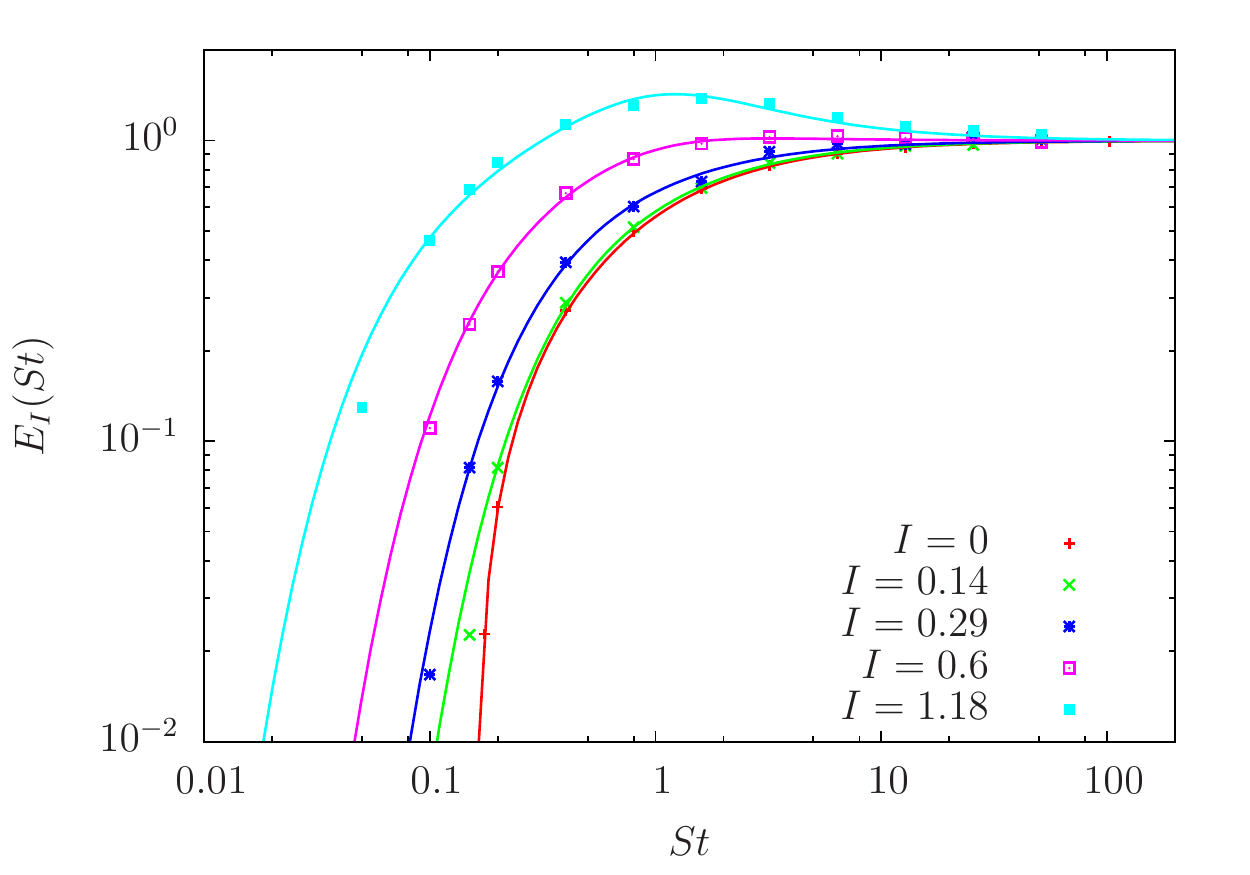}
\caption{\label{fig:collisionRate_lam_turb} Collision efficiencies for
  several turbulent intensities $I=\v_L/U_c$ \MODHOL{and
    $\Rep=400$. The $I=0$ curve is the same as in
    Fig.~\ref{fig:collisionRate_comp}}. Solid lines: fits of the
  function $E_I(\St)=\exp(-a(I)/\St)\,(1+b(I)\,\St/(1+\St^2))$ with $a=0.49$,
  $b=-0.095$ for $I=0.14$; $a=0.38$, $b=-0.04$ for $I=0.29$; $a=0.21$,
  $b=0.31$ for $I=0.60$, $a=0.085$, $b = 1.08$ for $I=1.18$.}
\end{figure}

Turbulent fluctuations significantly increase the collision
probability. Figure~\ref{fig:collisionRate_lam_turb} shows the
collision efficiency $E_I$ for various $I$. One observes that higher
is the turbulent intensity $I$, more collisions happen. This increase
is the strongest at small $\St$ and disappears asymptotically at large
$\St$.  As is discussed \MODHOL{later}, the collision efficiency
around $\St\approx 1$ remarkably exceeds unity.

\begin{figure}[h]
\centering
\includegraphics[width=0.9\hsize]{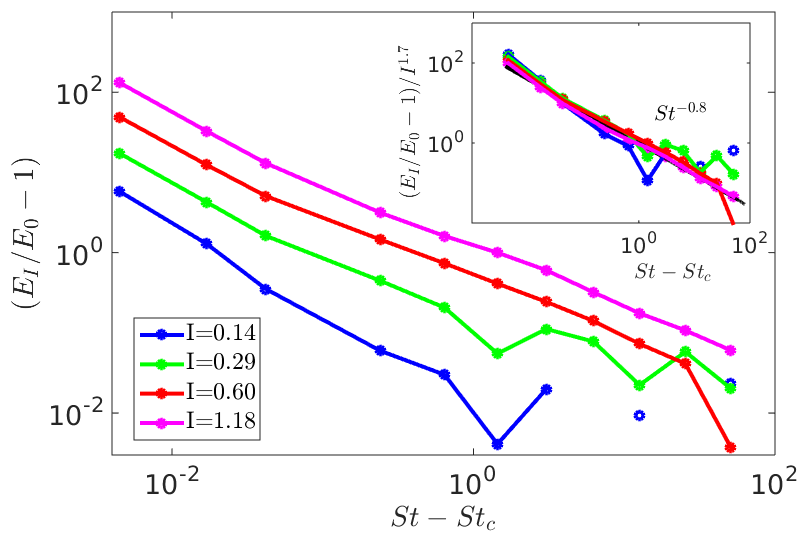}
\caption{\label{fig:collisionRateInc} Relative increase of the
  collision efficiency for several turbulent intensities as a function
  of the distance from the critical Stokes number of the laminar
  case.}
\end{figure}

For the smallest Stokes numbers and the largest turbulent intensity we
observe more than one hundred times more collisions than in the
reference laminar flow. This relative increase $\Delta
E(\St,I)=E_I/E_0-1$ of the turbulent efficiency compared with the
laminar one is shown in Fig.~\ref{fig:collisionRateInc}. It is
represented as a function of $\St-\St_c$ that is the distance from the
critical Stokes number $\St_c$ as $\Delta E$ diverges when
$\St\to\St_c$.  The \MODHOL{large values of $\Delta E$} at small $\St$
decreases as a power law with an exponent close to $-1$. The
dependence of $\Delta E$ on $I$ is nearly quadratic. Indeed, the
different curves almost fall on the top of each other when normalized
by $I^{1.7}$ \MODHOL{as can be seen from the inset of
  Fig.~\ref{fig:collisionRateInc}}.

\begin{figure}[h]
\centering
\includegraphics[width=0.9\hsize]{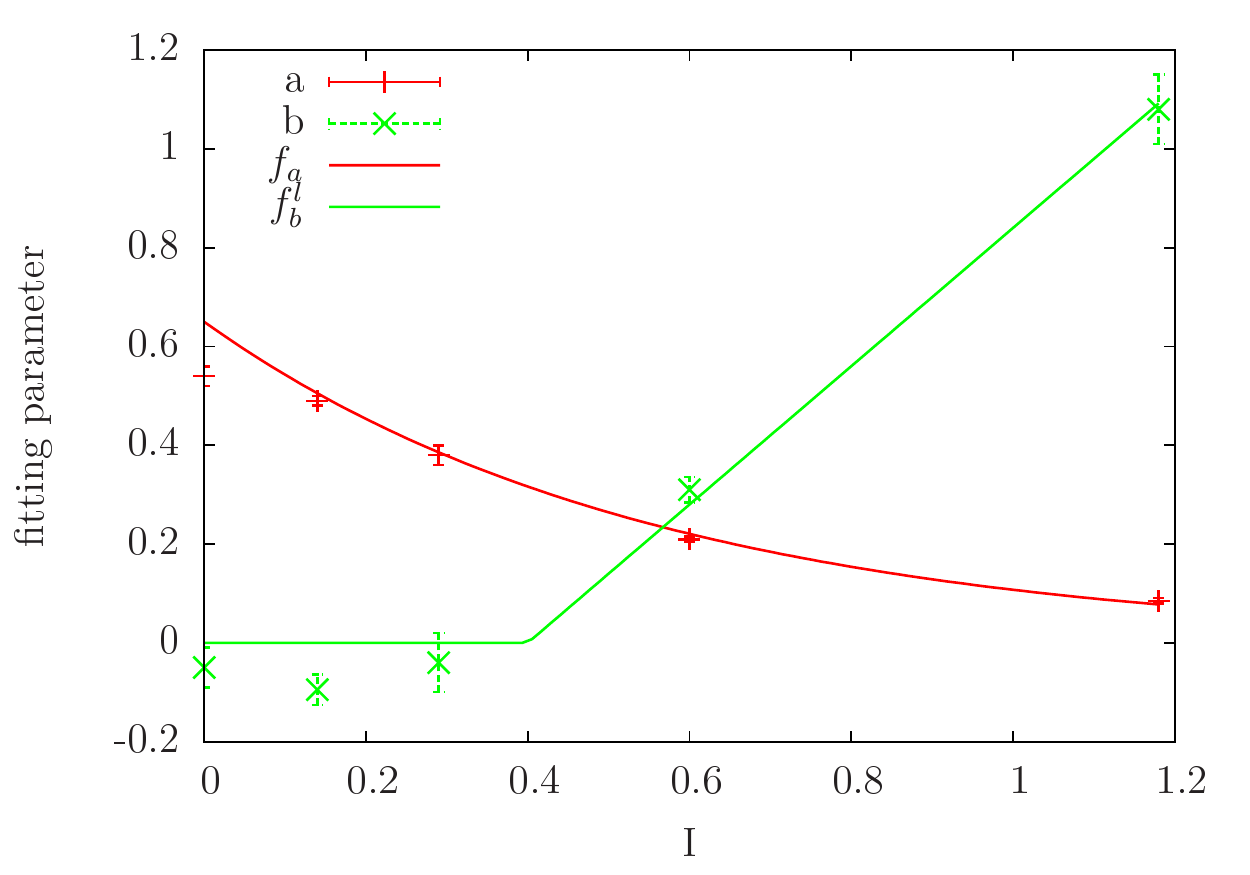}
\caption{\label{fig:fittingParameterE} Dependence of the fitting
  parameter $a$ and $b$ on the turbulent intensity $I$, for
  $\Rep=400$. $f_a=0.65\,\exp(-1.8\,I)$ is fitting function for the
  parameter $a$, $f^l_b=1.4\,(I-0.4)$ (for $I>0.4$) is a linear fit
  for the parameter $b$. \MODHOL{Both fits and the corresponding error
    bars were obtained by a standard least square method.}}
\end{figure}

Turbulent collision efficiencies (compare
Fig.~\ref{fig:collisionRate_lam_turb}) are well fitted by the function
\begin{equation}
\label{eq:collisionRate_fit}
E_I(\St)=\exp(-a(I)/\St)\,(1+b(I)\,\St/(1+\St^2))
\end{equation}
containing two parameters $a$ and $b$.  The dependence of these
parameters on the turbulent intensity is shown in
Fig.~\ref{fig:fittingParameterE} together with simple fitting
formulas. The parameter $a$ is decreasing roughly exponentially as a
function of $I$. $b$ is close to zero for small intensities and
strongly increasing starting from $I\approx 0.4$.

\begin{figure}[h]
\centering
\includegraphics[width=0.9\hsize]{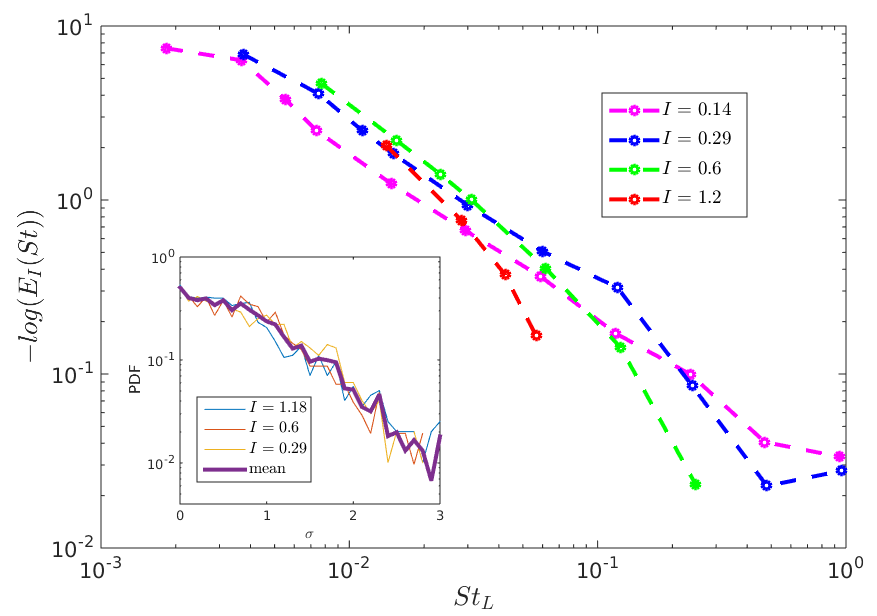}
\caption{\label{fig:collisionRate_log_gradient} Logarithm of collision
  efficiencies showing the exponential character at small
  $\St_{L}=\taus\,\v_{L}/L$. inset: Probability density function (PDF)
  of stream-wise velocity gradient \MODHOL{normalized to standard
    deviation} within the boundary layer.}
\end{figure}

In the small Stokes number limit, the turbulent collision efficiency
displays a clear exponential falloff (see
Fig.~\ref{fig:collisionRate_log_gradient}) indicating that the
critical Stokes number disappears when turbulence influences the dust
motion. Additionally, curves for different $I$ fall approximately on
top of each other once shown as a function of
$\St_{L}=\taus\,\v_{L}/L$. It is thus a turbulent time scale, namely
the large-eddy turn-over time, that replaces the advection time
$d/U_c$ relevant for the laminar problem.

\MODHOL{There are different ways in which turbulence enhances the
  collision probability. Fluctuations of the headwind speed mix
  collision efficiencies for different $I$ and $\St$ and allow for
  collisions of dust particles beyond the critical Stokes number
  $\St_c$. In turbulent gas flows, the dust heads onto the
  planetesimal with a fluctuating velocity. The probability
  distribution of the headwind is given by the one-point turbulent
  velocity distribution that is known to be close to a Gaussian. For
  the present problem, its standard deviation is given by the
  turbulent intensity $I$. These headwind variations lead to
  fluctuating planetesimal Reynolds number and dust Stokes
  numbers. Especially close to the laminar value $\St_c$ this results
  in higher collision probabilities and an absence of a critical
  Stokes number.}  Another consequence of a variable headwind are
fluctuations of the stream-wise velocity gradient $\sigma$ in the
upstream boundary layer of the planetesimal. They modify the local
Stokes number that can be defined by $\St=\taus\, \sigma$. The
probability for a collision of a dust particle with stopping time
$\taus$ is then $\mathbb{P}(\St>\St_c) =
\mathbb{P}(\sigma>\St_c/\taus)$, and thus relates to the probability
of observing a large velocity gradient at the particle surface.  The
probability density function (PDF) of $\sigma$ shown in the inset of
Fig.~\ref{fig:collisionRate_log_gradient}. \MODHOL{The tails are close
  to exponential, although we cannot rule out stretched exponential
  tails as observed in homogeneous isotropic turbulence. Exponential
  tails mean} $\mathbb{P}(\sigma>\St_c/\taus) \sim
\exp(-C\,\St_c/\taus)$ which is consistent with the formally observed
exponential fall off of $E_I(\St)$ at small values of $\St$.

\MODHOL{Another mechanism to increase the collision probability
  relates to turbulent diffusion. \cwosug{The later}{This} enhances the mobility of
  dust and increases its flux onto the planetesimal. As a simplified
  \textit{gedankenexperiment} one can think of a sphere moving in a sinusoidal
  velocity field representing the large-scale turbulent fluctuations.
  Let us assume (in the reference frame of the sphere) a velocity of
  the form ${\bm u}=U_c{\bm e}_x + I\,\sin(2\pi/t_L,t)\,{\bm
    e}_y$. The volume swept by the sphere is evidently larger in the
  turbulent case ($I>0$) than in the laminar ($I=0$). According to
  inertia, dust particles are more or less coupled to the gas which
  makes the swept volume additionally depending on $\St$. Small $\St$
  particles stick to the gas trajectories so their swept volumes
  equal. This effect can explain the observation of collision
  efficiencies exceeding unity in turbulent flows. For large $\St$,
  particles move ballistically so that they only sweep through the
  laminar ($I=0$) volume.  }

\begin{figure}[h]
\centering
\includegraphics[width=0.9\hsize]{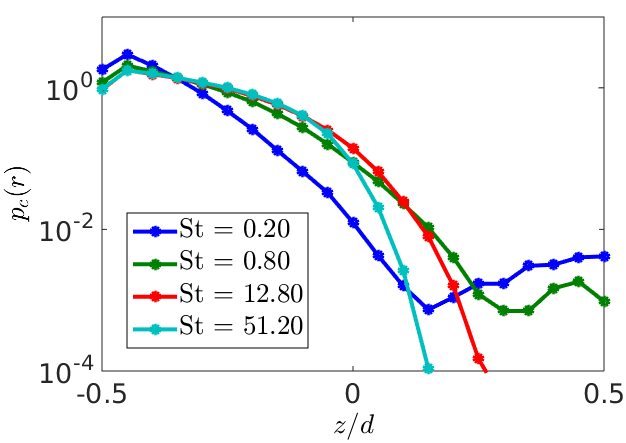}
\includegraphics[width=0.9\hsize]{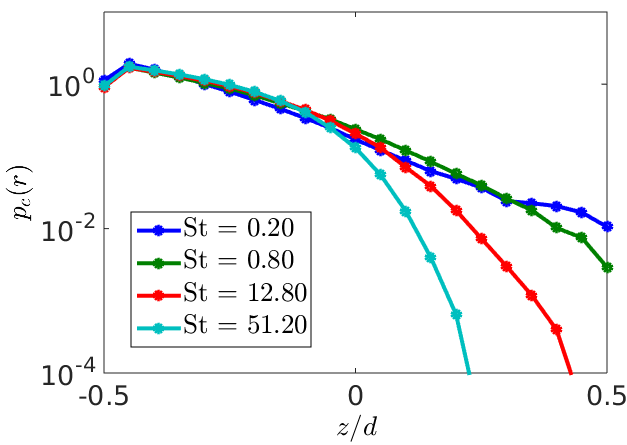}
\caption{\label{fig:p_stream} Probability of collisions as a
  function of the stream-wise position for $I=0.29$ (top) and $I=0.6$
  (bottom).}
\end{figure}

We conclude this section with a study of the spatial distribution of
dust impacts on the surface of a planetesimal in a turbulent disk. In
Fig.~\ref{fig:accretionPosition_r} we saw that dust collisions
preferentially happen close to the stagnation point of the flow in a
quiescent disk. This is still true when turbulent fluctuation agitate
the dust (see Fig.~\ref{fig:p_stream}). But the added randomness leads
to a homogenization of the impact position. The larger is $I$ the more
the collisions fill the entire planetesimal surface. And especially
for small $\St$ particles, backward collision become frequent.

\subsubsection{Impact velocity}

The relative velocity of dust and a planetesimal at impact is crucial
for the dust accretion problem as it determines, together with the
angle of impact, the outcome of a collision. Low collision speeds lead
to sticking of dust on the target surface, while high speeds lead to
bouncing, fragmentation with mass transfer or erosion
\citep[e.g.,][]{Blum+Wurm2008,windmark-birnstiel-etal:2012}.

\begin{figure}[h]
\centering
\includegraphics[width=0.9\hsize]{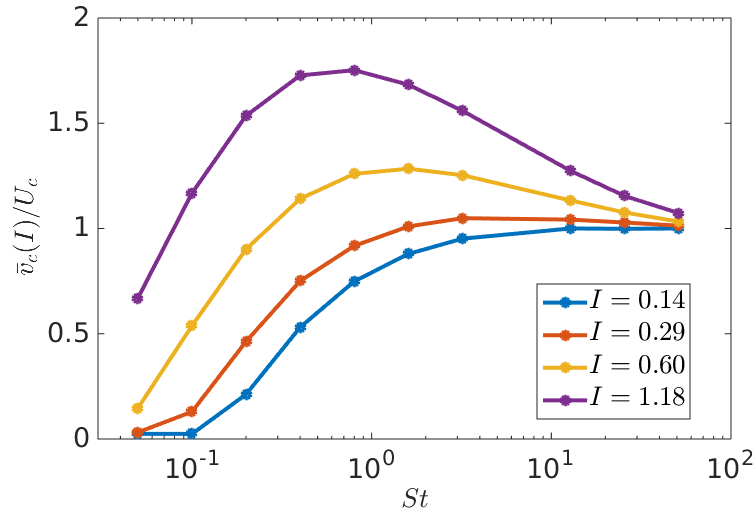}
\caption{\label{fig:vMean}Mean dust velocity at impact on the
  planetesimal surface for $Re=400$.}
\end{figure}

In laminar disks the mean impact speed of small-size dust (with a
stopping time smaller than the orbital period) is a monotonously
increasing function of inertia \citep{Weidenschilling1977}. Dust
particles with inertia close to the critical Stokes number only
slightly touch the planetesimal surface while large $\St$ particles
collide with the full headwind speed. Dust particles in turbulent
disks experience gas velocity fluctuations and in turn drag
variations. They follow preferentially turbulent structures with
characteristic time-scales that equal their response time. This
coupling of inertial particles is known to create non-trivial
phenomena such as the mentioned small-scale preferential
concentrations that are the most effective for
$\St_\mathrm{Kol}=\taus/t_\mathrm{Kol}\approx 0.6$ particles
\citep{reade2000effect,bec-etal:2006}.

We observe such a \MODHOL{\cwosug{eddie}{eddy}-dust coupling} also for
the collision velocity \MODHOL{$v_c$ (the norm of the dust velocity
  vector at impact)} of dust particles with a planetesimal (see
Fig.~\ref{fig:vMean}). Asymptotically, small-$\St$ dust (small in the
sense of particles with a small collision efficiency) still only
mildly touches the surface, while large-$\St$ dust collides with the
speed of the headwind. However, at intermediate values of $\St$,
turbulent velocity fluctuations lead to an increase of the collision
speed that even exceeds the headwind speed. For the highest turbulent
intensity that is studied here ($I=1.18$), the average impact speed is
approximately 75\% higher than the mean headwind speed. We remark that
once particle inertia is measured in terms of the characteristic time
scale of turbulent structures of size $d$ (planetesimal diameter)
$t_d=t_L\,(d/L)^{2/3}$ ($t_d=4.4, 2.1, 1.1$ for $I=0.29, 0.6, 1.18$)
all maxima of the curves (located at $St\approx 3.2, 1.6, 0.8$) in
Fig.~\ref{fig:vMean} align at $\St=\taus/t_d\approx 0.6$. Turbulent
eddies of the size of the planetesimal are thus responsible for this
increase of the impact velocity.

\begin{figure}[h]
\centering
\includegraphics[width=0.9\hsize]{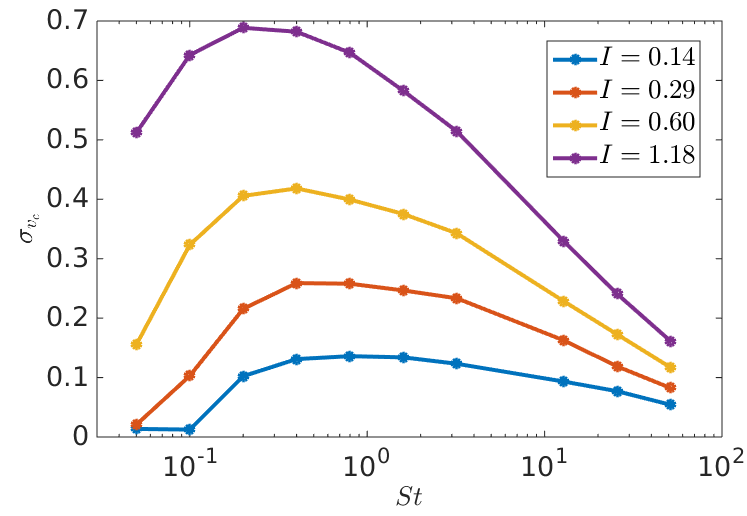}
\caption{\label{fig:vStd}Standard deviation of the impact speed for
  $Re=400$ and several turbulent intensities.}
\end{figure}
To estimate the broadness of the velocity distributions we measured in
Fig.~\ref{fig:vStd} their standard deviation. Here, the $\St\approx 1$
\MODHOL{peak} is even more important than for the mean impact
speed. For a $\St=0.4$ dust particle in a $I=1.18$ headwind, the
velocity distribution is up to six times broader than in a
non-turbulent disk.

\begin{figure}[h]
\centering
\includegraphics[width=0.9\hsize]{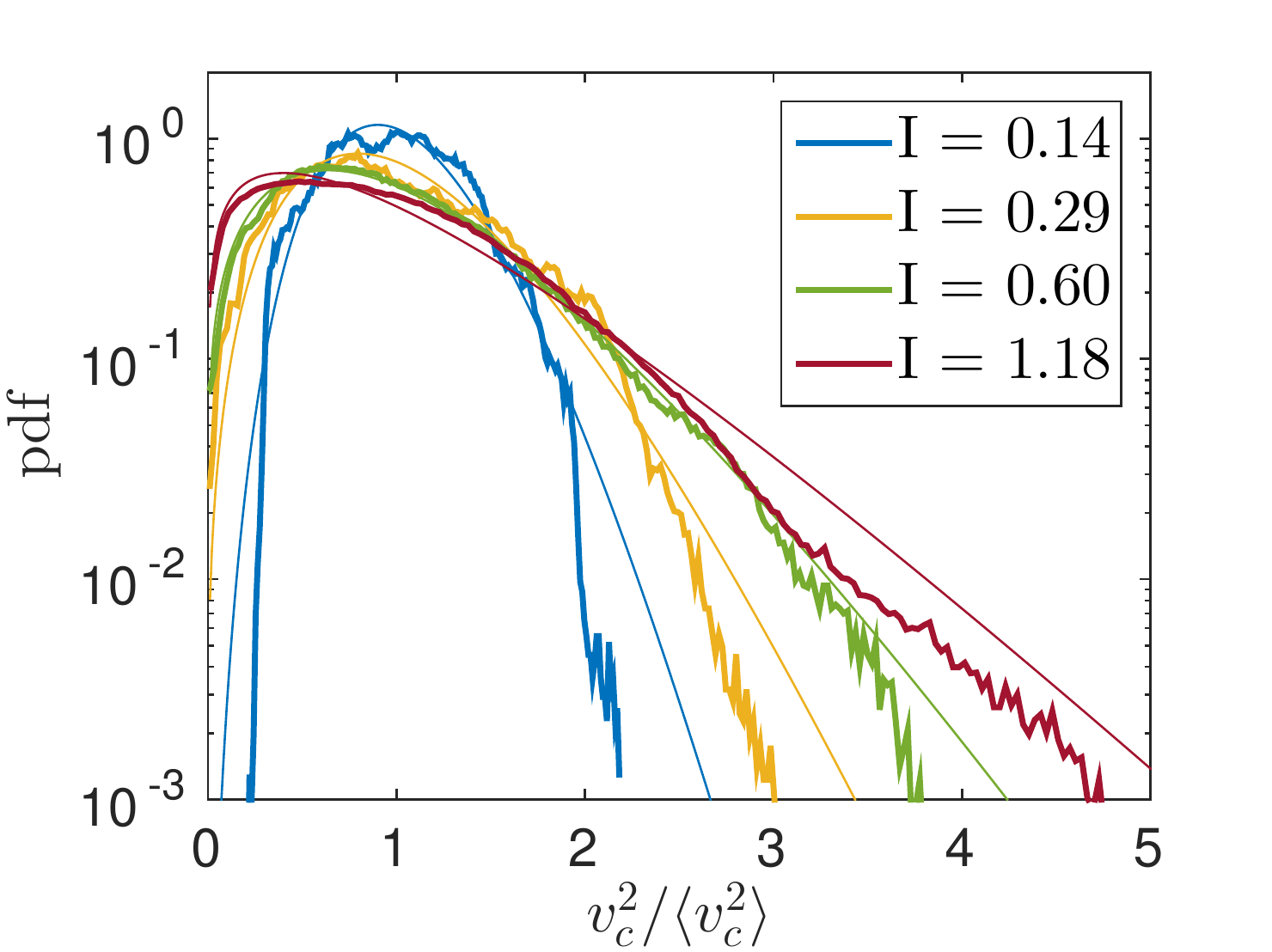} \hfill
\caption{\label{fig:vPerpDist_I} Probability density function of the
  impact velocity normal to the planetesimal surface for $\St=0.8$ and
  several turbulent intensities \MODHOL{as labeled. The bold lines
    correspond to measurements from numerical simulations, while the
    thin lines refer to the non-central chi-squared prediction (see
    text).}}
\end{figure}

\begin{figure}[h]
\centering
\includegraphics[width=0.9\hsize]{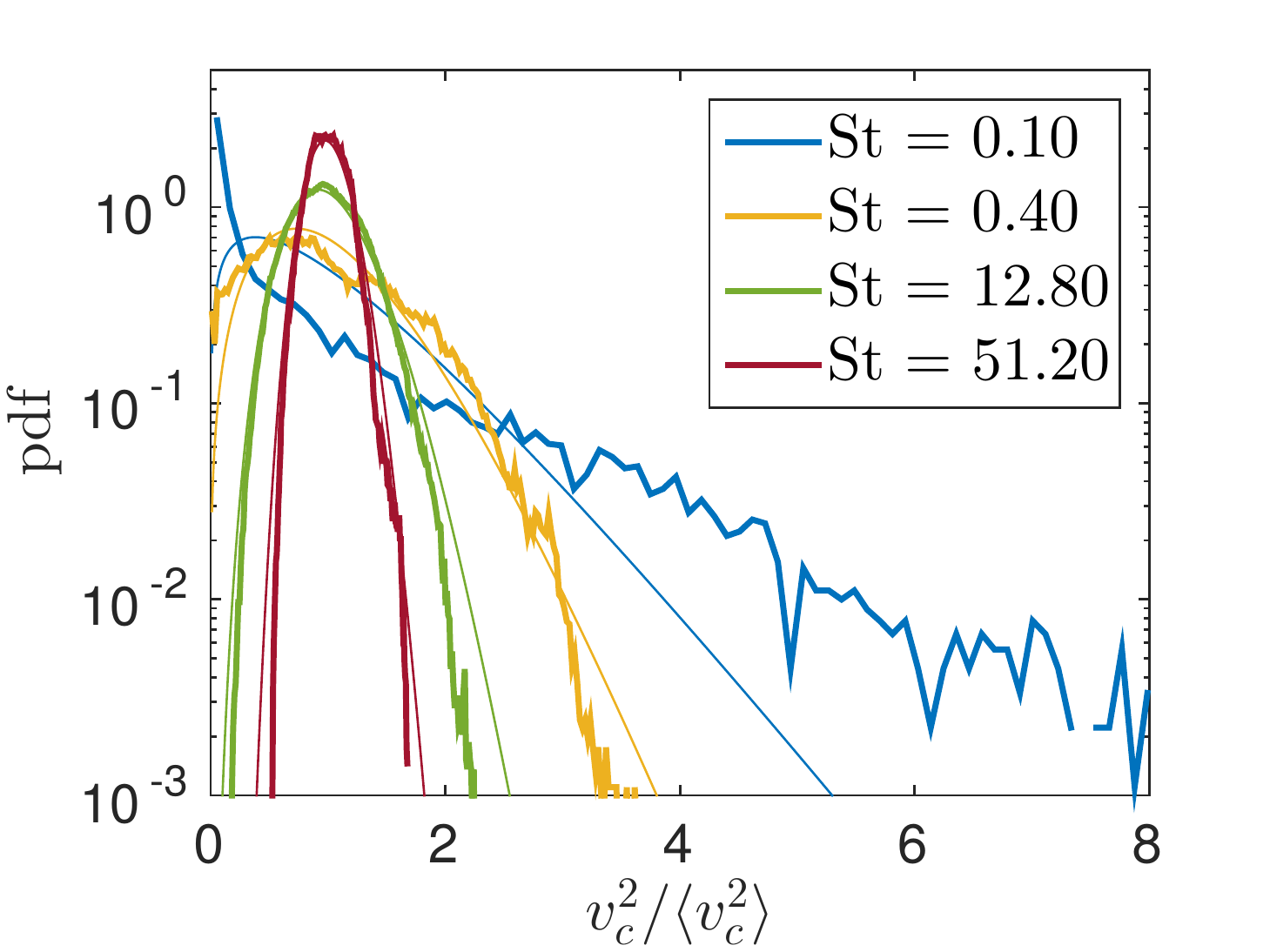}
\caption{\label{fig:vPerpDist_St} Probability density function of the
  surface normal impact velocity for $I=0.29$ and several Stokes
  numbers \MODHOL{as labeled. As in Fig.~\ref{fig:vPerpDist_I}, the
    bold lines correspond to numerical simulations and the thin lines
    refer to the non-central chi-squared prediction.}}
\end{figure}
The impact speed has, besides its average value and standard
deviation, a probability distribution that varies with both $I$ and
$\St$. It reveals that high-impact speeds of the order of several
times the headwind speed are quite probable. For a fixed value of
$\St$ the distribution becomes monotonously broader with increasing
$I$ (see Fig.~\ref{fig:vPerpDist_I}).  \MODHOL{The numerical data is
  there compared to the non-central chi-squared distribution that
  would be obtained if $v_c$ were the norm of a three-dimensional
  random Gaussian vector with prescribed mean and variance.  Up to
  statistical accuracy, it seems from Fig.~\ref{fig:vPerpDist_I} that
  such an approach gives a rather good description of actual
  fluctuations of the impact velocity. This approximation is however
  valid only if the Stokes number $\St$ is sufficiently large.
  Figure~\ref{fig:vPerpDist_St} indeed represents the same
  distributions for a fixed value of $I$ at varying $\St$. One
  observes deviations from the chi-squared prediction in both tails at
  the smallest value of the Stokes number. It seems nevertheless that
  the distribution still belongs to the same family and can be
  approximated by a chi-squared distribution with a smaller number of
  degrees of freedom.  Everything happens as at small $\St$, the dust
  velocity fluctuations with respect to the planetesimal were
  constrained in a space with dimension less than three.  }

\begin{figure}[h]
\centering
\includegraphics[width=0.9\hsize]{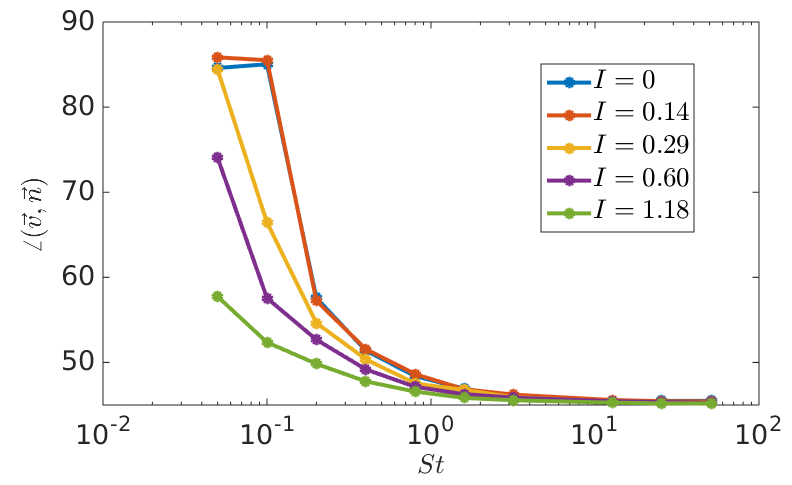}
\caption{\label{fig:angle}Average angle of impact
  $\bar\angle(\vec{v},\vec{n})$ with respect to the surface normal
  direction $\vec{n}$ in degrees.}
\end{figure}

The outcome of a collision also depends on the angle of
impact. Figure~\ref{fig:angle} shows the average collision angle
$\bar\angle$ with respect to the surface normal direction
$\vec{n}$. Small $\St$ dust preferentially touches the planetesimal
with an mean impact angle close to $90^0$. High inertia dust heads
straight onto the planetesimal and experiences an average impact angle
of $45^0$. Turbulent fluctuations randomize the impact angle and favor
this way to the same mean angle of
$\bar\angle(\vec{v},\vec{n})=45^0$. 

\section{Astrophysical application}
\label{sec:astro}

\subsection{Linear cross section}
We \cwosug{now want to} apply these results to \cwosug{real}{realistic} disk conditions. In order
to study filtering of dust by planetesimals, \citet{Guillot+2014} had
applied the results of numerical simulations by
\citet{SekiyaTakeda2003} for a laminar disk and a fixed planetesimal
Reynolds number $\Rep=50$. Our simulations in the laminar case cover a
range of $\Rep$ from $100$ to $1000$. As seen in Fig.~\ref{fig:mmsn}
and Eq.~\eqref{eq:Rep}, this corresponds to planetesimals with
diameters between 4\,m and 40\,m at 1\,au and between 1\,km and 12\,km
at 10\,au. Since the outcome of the simulations is only weakly
dependent on $\Rep$ (the critical Stokes number is a function of
$\log(1+\Rep/2)$ --see eq.~\eqref{eq:e_fit}), we expect to be able to
extrapolate the results outside this range.

In most cases however, turbulence is expected to be
important. \TG{Our simulations in the turbulent case have been
  calculated for various intensities of the turbulence $I$, but a
  fixed planetesimal Reynolds number $\Rep=400$. However, when
  turbulence becomes important,} we expect the results to become very weakly dependent on
$\Rep$. This is for two reasons: First as seen in
Fig.~\ref{fig:flow_particles}, for values of $I$ approaching unity,
the flow around planetesimals is perturbed very significantly \TG{and
becomes controlled by the turbulence of the disk instead of by the
planetesimal properties. Second, turbulence perturbs the boundary
layer around the planetesimal independently of its properties to offer new possibilities for dust
particles to impact}.

But for small planetesimals and/or \cwosug{weak turbulence}{low
  turbulent intensity}, the planetesimal size can become smaller than
the Kolmogorov scale, i.e., the minimum scale for turbulent eddies. In
that case, the planetesimals experience a headwind of variable
intensity and direction. It is expected that, in the limit of
$d/\lkol\ll 1$, the situation becomes similar to the laminar case, but
with a headwind that is increased by $\sqrt{1+I^2}$.  We write
\begin{equation}
\label{eq:fhydro}
f_{\rm hydro}=
\begin{cases}
  \left[E_0(\St^*,\Rep)\right]^{1/2} & \quad \mbox{if
    $\delta_\mathrm{Kol}=d/\ell_{\rm Kol}<1$}\\[.5em]
  \left[E_I(\St)\right]^{1/2} & \quad \mbox{otherwise}
\end{cases}
\end{equation}
\cwosug{and}{where $f_\mathrm{hydro}$ is the collision efficiency as defined by \citet{Guillot+2014}, }
$\St^*=\St\sqrt{1+I^2}$\cwosug{. We use for}{ and} $E_0$ and
$E_I$ the fitting formula (\ref{eq:e_fit})
and~(\ref{eq:collisionRate_fit}) \cwosug{that were proposed in previous
section}{}. With this definition, $2R_{\rm p}f_{\rm hydro}$ is the
planetesimal linear \cwoadd{collisional} cross section and $\pi R_{\rm p}^2f_{\rm hydro}^2$
its surface \cwoadd{collisional} cross section.  Thus, for a planetesimal smaller than the
smallest turbulent eddy, the flow is considered laminar, but we
account with the use of $\St^*$ for a flow velocity that is slightly
higher on average. On the other hand, when the planetesimal is larger
than the Kolmogorov scale, we use the results of the simulations in
the turbulent case directly.

\begin{figure}
\centering
\includegraphics[width=\hsize]{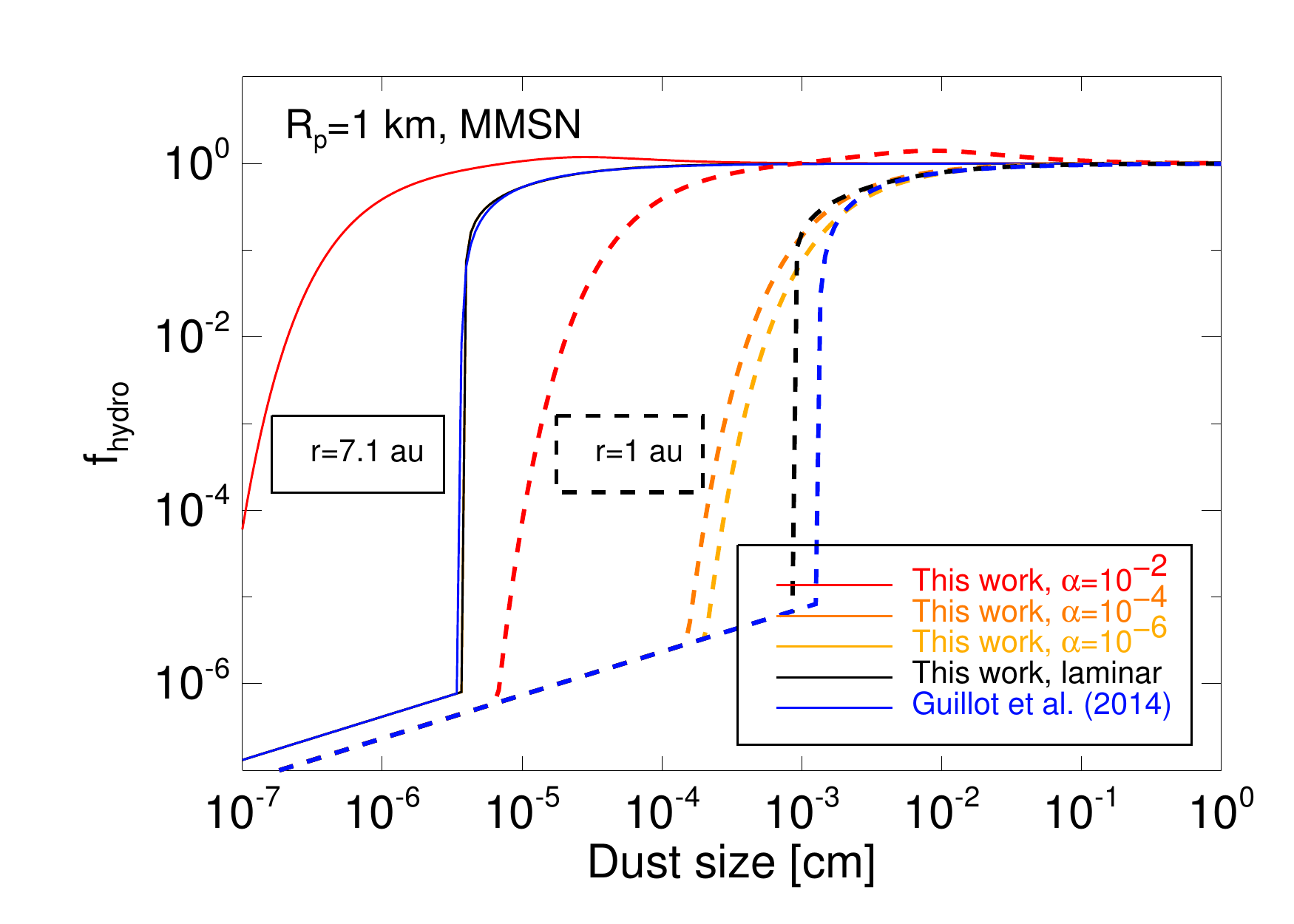}
\caption{\label{fig:fhydro}Value of the factor $f_{\rm hydro}$
   \TG{(see eq.~\eqref{eq:fhydro})} indicative of a reduction of the linear cross section of
  planetesimals resulting from hydrodynamical effects as a function of
  dust particle size, for a planetesimal of 1\,km radius in
  an MMSN disk, at an orbital distance of 7.1\,au (plain) and 1\,au (dashed) 
  and for various levels of turbulence (as labeled). The results are  
  compared to the one obtained by \citet{Guillot+2014} which is
  independent of turbulence amplitude. For the 7.1\,au case (which
  corresponds to $\Rep=400$ as in the previous hydrodynamical
  calculations), the lines
  corresponding to $\alpha=10^{-4}$ and $10^{-6}$ are hidden behind the
  laminar case (see text).}
\end{figure}

Figure~\ref{fig:fhydro} illustrates how the factor $f_{\rm hydro}$
varies as a function of particle size in an MMSN disk for a
1\,km-radius planetesimal either at 7.1\,au or at 1\,au. The first
case corresponds to a planetesimal Reynolds number $\Rep=400$ equal to
the one used in the hydrodynamical simulations with turbulence. The
second case corresponds to a much higher Reynolds number
$\Rep=5.4\times 10^4$ outside the range of our simulations.

In all cases, particles which are larger than a critical value (i.e., with a
Stokes number higher than unity) are accreted with a cross-section
approximately equal to the geometric one (i.e., $f_{\rm hydro}\approx
1$). In laminar disks or when turbulence is small, the cross section
drops for small particles such that $\St<1$. If turbulence is large
enough, this drop occurs at even smaller sizes, with an offset that
corresponds to one to two orders of magnitude for the case with
$\alpha=10^{-2}$. 

The comparison of the laminar $\Rep=400$ cases shows a relatively good
agreement between our work and the previous results of
\citet{Guillot+2014} who used results from \citet{SekiyaTakeda2003}
for a fixed planetesimal Reynolds number of $50$. When turbulence is
added, it is worth noticing that while the case with $\alpha=10^{-2}$
stands out and allows much smaller particles to collide, the cases
with $\alpha=10^{-4}$ and $10^{-6}$ are almost indistinguishable from
the laminar case. This is a direct consequence from the fact that
$\delta_\Kol<1$ for these: the smallest turbulent cell is expected to
be larger than the planetesimal size which implies that we switch to
the laminar case in eq.~\eqref{eq:fhydro}. Our approach is thus
discontinuous in $\alpha$, but resolving this issue would require
dedicated simulations beyond the scope of the present work

At high Reynolds number (i.e., the 1\,au case in
Fig.~\ref{fig:fhydro}), the Kolmogorov parameter $\delta_\Kol$ is
generally high which implies a nearly continuous behavior from high to
low values of $\alpha$. A small issue seen for low values of the
viscosity and dust sizes corresponding to Stokes number close to unity
is that the value of $f_{\rm hydro}$ for a disk with low turbulence
(e.g., $\alpha=10^{-6}$) can become smaller than the laminar value,
which according to our simulations is unlikely. Clearly, this is a
consequence of the fact that our expressions have been derived for a
relatively low planetesimal Reynolds number and are applied very far
from that value. Again, dedicated simulations would be needed, but
\cwosug{with the additional complication that they would require prohibitively high computing power.}{may be out of reach with present-day computing power.}

\subsection{Collision probabilities in disks}

We now examine the consequences for collisions of dust grains with
planetesimals with the same approach as \citet{Guillot+2014}. 
\TG{%
In protoplanetary disks, collisions between drifting dust particles
and planetesimals occur with a probability ${\cal P}_{\rm 3D}$ that is
a function of the planetesimal cross section, the scale height of the
dust disk $h_{\rm d}$ and of the drift velocity of the dust
particles. The latter depends on orbital distance $r$, orbital
(keplerian) frequency $\Omega$ and stopping time $t_{\rm s}$. In the
limit that gravitational effects and gas drift may be neglected and
for circular orbits, this probability can be shown to write
\citep{Guillot+2014}: 
\begin{equation}
{\cal P}_{\rm 3D}={1\over 2\sqrt{\pi}}{R_{\rm p}^2 f_{\rm hydro}^2\over H_{\rm d} r}\sqrt{1+{1\over 4 t_{\rm s}\Omega}},
\end{equation}
where $f_{\rm hydro}$ accounts for hydrodynamical effects discussed previously (the purely geometrical limit is recovered for $f_{\rm hydro}=1$). }

\TG{In reality, gravity becomes important both for median to large planetesimals (kilometer size and more) and for large grains (above meter size) and seriously complicates the picture. Several interaction regimes may be defined as follows} \citep[see][]{ormel2010effect, Guillot+2014}: 
\begin{itemize}
\item The geometric regime,
corresponds to the most simple case in which drag, hydrodynamical
and gravity effects may be neglected. 
\item\TG{We define the hydrodynamical regime 
as an extension of this regime at small dust sizes when we must
account for the deflection  of dust grains around planetesimals.} 
\item The Safronov regime corresponds
to the case when large ``dust'' (effectively, boulders) which are very
weakly affected by gas drag migrate inward so slowly that they feel a
gravitational focusing by the planetesimal which increases the
collision probability. 
\item In the three-body regime, the gravity fields of the
planetesimal and that of the central star must be taken into
account. 
\item The settling regime corresponds to the case when
gravitational acceleration from the planetesimal and gas drag on the
dust particles lead to an enhanced capture probability. 
\end{itemize}

\begin{figure}
\centering
\includegraphics[width=\hsize]{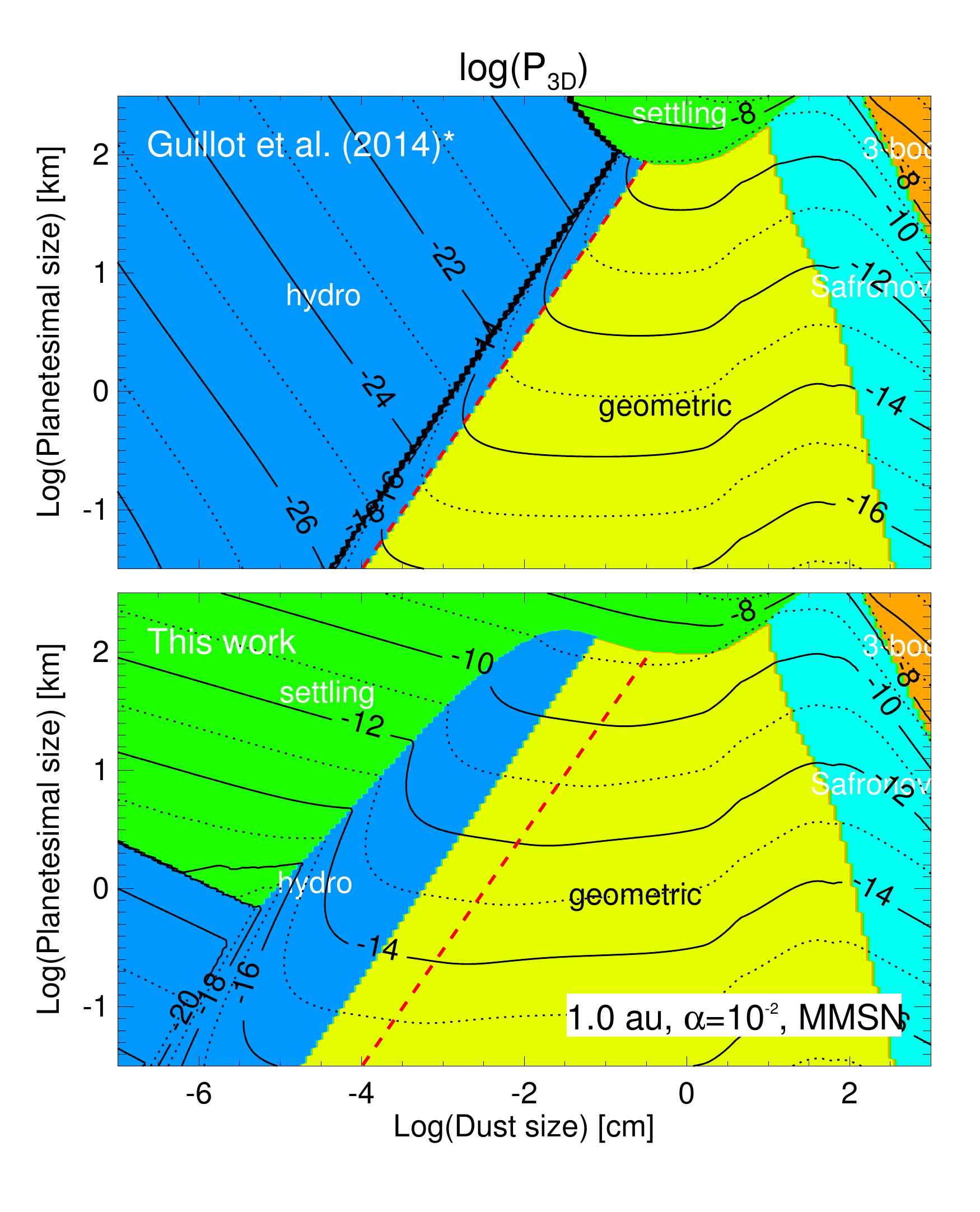}
\caption{\label{fig:prob_fixedrorb}Contours of the collision
  probability ${\cal P}_{\rm 3D}$ obtained at 1\,au in an MMSN disk
  with $\alpha=10^{-2}$ as a function of dust size and planetesimal
  size. The colored areas (labeled ``geometric'', ``hydro'',
  ``Safronov'', ``three-body'' and ``settling'') correspond to
  different interaction regimes. The top panel shows the results
  obtained using the same approach as \citet{Guillot+2014} but
  corrected for a factor $f_{\rm hydro}$ (see text and compare with
  their Fig.~16). The bottom panel corresponds to results with the new
  prescriptions for the hydro model. For an easier comparison, the
  dashed red line marks the location of the hydro regime
  (corresponding to $f_{\rm hydro}<0.9$) in the \citet{Guillot+2014}
  study. }
\end{figure}

Accounting for the complexity of the problem we thus calculate the
collision probability between dust and planetesimals, ${\cal P}_{\rm 3D}$,
from eq.~(43) of \citet{Guillot+2014}, assuming monodisperse size
distributions\footnote{In doing so, we correct for the fact that in
  \cite{Guillot+2014}, the 3D collision probability in the hydro mode
  was overestimated because it neglected the reduction in the vertical
  cross section, i.e., ${\cal P}_{\rm 3D}\propto f_{\rm hydro} R_{\rm p}^2$
  had been assumed instead of
  ${\cal P}_{\rm 3D}\propto f_{\rm hydro}^2 R_{\rm p}^2$. Because this
  affected the hydro mode with an already very low collision
  probability this had negligible effect on the qualitative results.},
\TG{but including $f_{\rm hydro}$ from eq.~\eqref{eq:fhydro}. In
doing so, we also adopt an important modification stemming from the
work of \cite{Johansen+2015} and \cite{Visser+Ormel2015}: Instead of limiting the
extent of the settling regime to when the capture radius is larger
than the physical size of the planetesimal as in \cite{Guillot+2014},
we instead look for solutions of the settling regime equations outside
of this range and adopt for the collision probability the maximum of
the probabilities obtained in the settling and geometric+hydro
regimes. This is important in regions where $f_{\rm hydro}$ is
extremely low but gravity and gas drag can still affect the trajectories of the
dust particles. }

Figure~\ref{fig:prob_fixedrorb} shows how ${\cal P}_{\rm 3D}$ varies with
dust and planetesimal size for a fixed orbital distance of
1\,au. We focus on planetesimals smaller than 100\,km and down to 
10\,m with the caveat that for planetesimals smaller
than about 1\,km, gas drag should be included. \TG{The top panel shows
the previous results from \cite{Guillot+2014}, which correspond to the
case of a laminar flow and no extension of the settling regime. The
bottom panel shows the results for a turbulent flow with full account
for gravity effects even for low-planetesimal sizes.}

\TG{The comparison between the top and bottom panels of
  fig.~\ref{fig:prob_fixedrorb} shows that even a weak planetesimal
  gravity effectively limits the decrease of the collision
  probabilities in the extended settling regime for dust smaller than
  $\sim 100\,\mu$m and planetesimals between one and 100\,km. The
  inclusion of turbulence effects also } shifts the hydrodynamic
regime to smaller dust sizes. The shift is about one order of
magnitude for all planetesimal sizes considered when comparing the
results for $\alpha=10^{-2}$ to those for a laminar disk. Particles of
0.1\,mm can hence be accreted relatively efficiently by planetesimals
for all the sizes considered. However, smaller particles still end in
the hydrodynamical regime with a strongly reduced collision
efficiency. For example micron-sized dust particles are very
inefficiently captured by planetesimals larger than a few kilometers
in size.

\subsection{\TG{The inefficient capture of small dust grains}}

\begin{figure}
\centering
\includegraphics[width=\hsize]{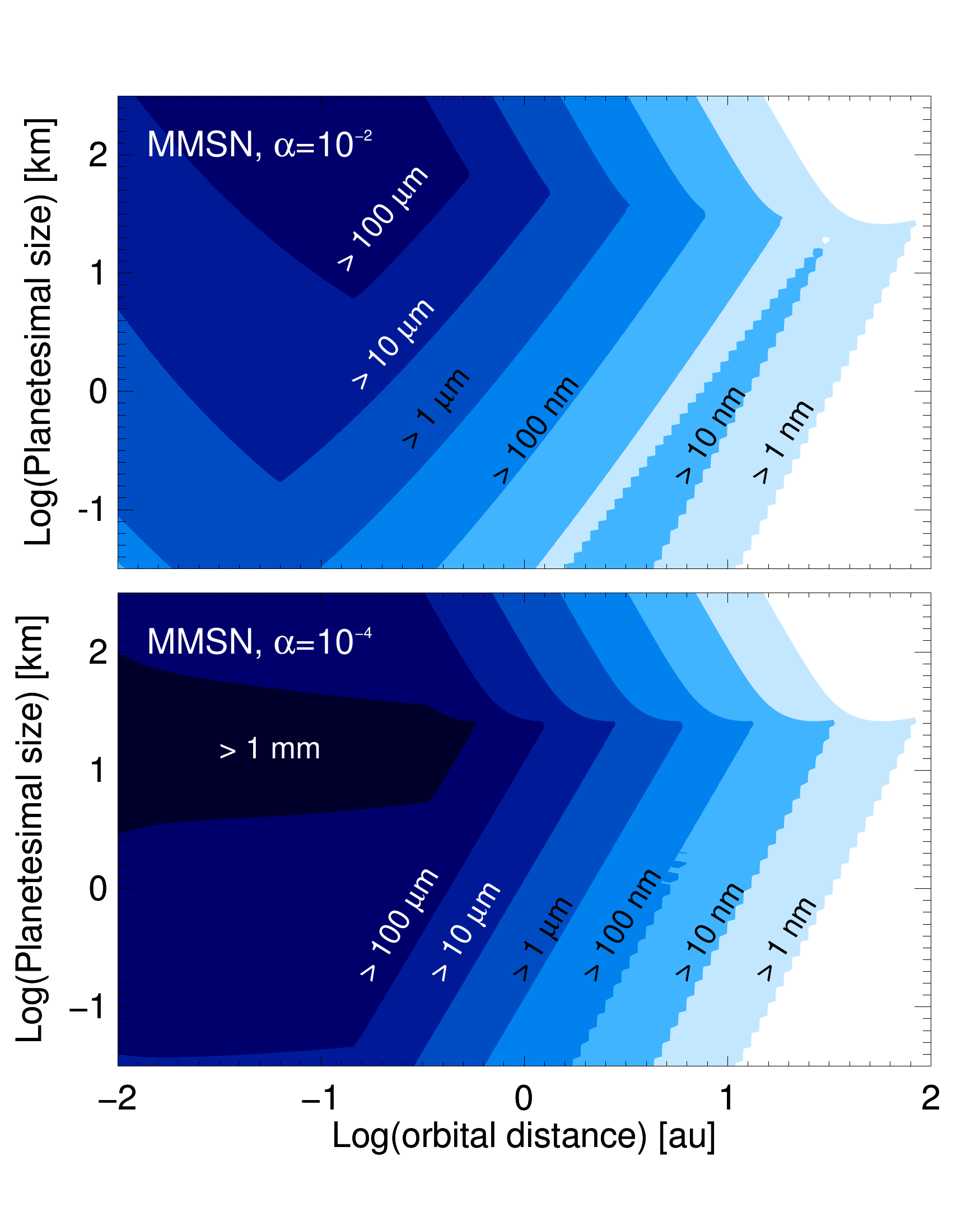}
\caption{\label{fig:prob_fixeds}\TG{Regions in which the collection of
  dust grains becomes very inefficient (defined as a collision cross section lower than 1\% of
  the geometric one --see text)} as a function of orbital distance and
  planetesimal radius for various 
  dust sizes (labeled) and for two values of the turbulence:
  $\alpha=10^{-2}$ (top panel) and $\alpha=10^{-4}$ (bottom
  panel). \TG{Large particles are collected relatively efficiently
  everywhere except in the innermost regions of the disk where the gas
  density is assumed to be high. Small particles are inefficiently
  collected everywhere except at the largest orbital distances, at
  least in the case of a disk characterized by the MMSN gas density
  distribution.} }
\end{figure}

\TG{We now turn to the examination of how (in)efficiently individual small
  dust grains may have been collected by planetesimals as a function
  of their sizes and orbital distance. A full model
  would require studying the size distribution of dust and
  planetesimals and is beyond the scope of the present paper. However,
  we can identify the parameter space for which this collection of
  dust is inefficient by identifying when the collision cross section
  becomes smaller than $1\%$ of the geometrical one (i.e.,
  corresponding to $f_{\rm hydro}<0.1$ in the limit when gravity
  effects are not important). Because the drop in collision
  probability in the hydro region of fig.~\ref{fig:prob_fixedrorb} is
  quite abrupt, we expect that if dust is indeed collected individually by planetesimals,
  this process should leave its imprint on the size distribution of
  individual grains in meteorites.}

\TG{Figure~\ref{fig:prob_fixeds} identifies these regions} as a function of
orbital distance and planetesimal size, \TG{either in the case of a high
turbulence level (top panel) or a low turbulence level (bottom
panel). In both cases, the collection of very small particles
(nanometer sizes) is found to be very inefficient, at least inside of
10\,au. Dust particles of progressively larger sizes can be collected
up to shorter orbital distances, but the efficiency then strongly
depends on the turbulence level.}

For 1\,mm particles (corresponding to a typical size of chondrules),
\TG{we do not see in fig.~\ref{fig:prob_fixeds} a region of strongly
  inefficient collection when turbulence is high ($\alpha=10^{-2}$),
  but} for $\alpha=10^{-4}$, these particles avoid
collisions with planetesimals \TG{between about 0.3 and 30\,km} within a fraction of
an au. \TG{One-micron particles are collected inefficiently inside a region
extending from about 0.1 to 3\,au depending on planetesimal size for
$\alpha=10^{-2}$, but this region extends from 0.8 to 6\,au for
$\alpha=10^{-4}$.} Smaller particles can collide with planetesimals only at larger
orbital distances, when the gas density has decreased and the stopping
time increased for a given particle size.

A larger turbulence level can therefore allow collisions of small-size
particles which would otherwise be avoided due to the hydrodynamical
flow around the planetesimals. However, this is also balanced by the
fact that higher turbulence means a thicker dust subdisk which lowers
the collision probability \citep[see][]{Guillot+2014}. \TG{Due to the form
of eq.~\ref{eq:fhydro}, the effect of
turbulence becomes} weaker at large orbital distances,
when the size of the smallest turbulent eddies becomes larger than the
planetesimals. This thus explains why the contour lines for very small
dust particles are identical for the $\alpha=10^{-2}$ and
$\alpha=10^{-4}$ cases.

The change in behavior of the contour plots for $\alpha=10^{-2}$, dust
sizes between 1\,$\mu$m and $1$ mm and orbital distances from 0.05 to
0.3\,au is due to a change in drag behavior for these particles: At
short orbital distances, the gas density is so high that they are in
the Stokes regimes and they switch to an Epstein drag beyond about
1\,au.

For the planetesimal sizes considered, particles smaller than about
10\,nm have a collision probability that is independent of alpha. This
is because collisions can occur only in the outer disk where
$\delta_{\rm Kol}<1$, i.e., the
smallest Kolmogorov scale is still larger than the planetesimals
considered. 

Small particles such as the presolar grains present in
meteorites, which can have sizes of only a few nanometers
\citep[e.g.,][]{ClaytonNittler2004} must have 
either collided with planetesimals far out in the disk or be incorporated
into larger grains which would have themselves collided with
planetesimals \TG{\citep[e.g.,][]{Ormel+2008}. For some of the
  presolar grains, given their very low abundance, it remains possible that they were
incorporated directly in planetesimals, although this would have to be
quantified}. In any case, this should have occurred without leading to any
melting or dissociation of these grains which kept their identify
throughout.



\section{Conclusions}
\label{sec:conclusions}

We have derived the accretion probability of small particles by a
planetesimal in a turbulent gas. In order to do so, we
performed high-resolution hydrodynamical simulations of the flow around a
spherical planetesimal of diameter $d$ moving with a velocity
$U_c$, assuming incompressibility. We studied both the case of a
laminar flow and that of a turbulent one, the intensity of the
turbulence being related to the turbulent viscosity of the disk. Dust
particles of variable size were implemented in the flow to determine
collision rates. 

For laminar flows, we confirm that small particles with a Stokes
number $\St<1$ (corresponding to stopping times shorter than the time
to cross the planetesimal) see a significant drop in their collision
rate with the planetesimal. For turbulent flows however, this drop
occurs for sizes that can be significantly smaller, i.e., turbulence
helps accreting dust particles with sizes up to one to two order of
magnitudes smaller than for laminar disks. 

We thus derived collision probabilities both in the laminar case
[eq.~\eqref{eq:e_fit}] and in the turbulent case
[eq.~\eqref{eq:collisionRate_fit}]. These expressions, even if limited
to limited to $\Rep=400$, can be used for a wide range of
situations. We propose an approximate recipe to use either the laminar
case if the planetesimal size is smaller than the Kolmogorov scale and
the turbulent case otherwise [eq.~\eqref{eq:fhydro}].

When applied to real disks, our new expressions \cwosug{lead to a
  shrinkage of the hydro region to small grain sizes}{shift the
  boundary with the hydro regime -- where accretion rates are greatly
  suppressed -- to smaller sizes}. For example, for $\alpha =
10^{-2}$, the upper limit dust size in the hydrodynamical regime is
decreased by a factor 100 and even sub-$\mu$m size particles collide
efficiently with one-kilometer planetesimals. They also show that the
accretion of extremely small particles is difficult and generally
requires to be done by small planetesimals (less than km size) at
large orbital distances (beyond 1\,au) and/or late in time, when the
disk has become less massive. We believe that these results are
important to interpret, among other things, the presence and
characteristics of presolar grains in meteorites since these vary in
size from several microns down to only a few nanometers.

\TG{In order to apply our results to protoplanetary disks, we had to
  approximate the effect of gravity, often by extrapolations far from
  the regime in which numerical experiments were conducted. Future
  efforts will be directed towards including the gravitational force
  directly in our hydrodynamical simulations.}

\appendix
\section{Definition of symbols}
\label{appendix}
\TG{We summarize here the symbols used in this paper and their definitions.}

\begin{table*}
  \caption{\label{table3} \TG{Symbols used in this article}} \centering
 \begin{tabular}{ccc}
   \hline\hline
   Symbol                      & Description                               & Equation \\
   \hline 
   $U_c$                       & average headwind velocity\\
   $I$                         & turbulent intensity                        & $\v_{L}/U_c$\\
   $Re_p$                      & planetesimal Reynolds number               & $U_c\,d/\nu_\mathrm{mol}$\\
   $Re$                        & outer gas flow Reynolds number              & $\v_{L}L/ \nu_\mathrm{mol}$  \\
   $\v_{L}$                     & root-mean-square velocity of the gas flow  & $\sqrt{2/3 E_k}$\\
   ${\bm u}$                   & gas velocity                               & \\
   $E_k$                       & kinetic energy of the gas flow              & $\frac{1}{2}\int |{\bm u}|^2$\\
   $\varepsilon_\mathrm{Kol}$    & mean kinetic energy dissipation rate       & $\frac{1}{2}\int |\nabla \times {\bm u}|^2 $\\
   $\nu_\mathrm{mol}$            & kinematic viscosity\\
   $d$                         & planetesimal diameter\\
   $R_{\rm p}$                         & planetesimal radius & $d/2$\\
   $\lkol$                     & Kolmogorov length scale                    & $(\nu_\mathrm{mol}^3/\varepsilon_\mathrm{Kol})^{1/4}$\\
   $t_\mathrm{Kol}$             & Kolmogorov time scale                       & $(\nu_\mathrm{mol}/\varepsilon_\mathrm{Kol})^{1/2}$\\
   $L$                        & integral scale                              & $\v_L^{3}/\varepsilon_\mathrm{Kol}$\\
   $t_L$                       & integral time scale                        & $L/\v_L$\\
   $t_s$                       & response or stopping time of dust          & \\
   $St$                        & Stokes number                              & $t_s/t_c$\\
   $St_c$                      & critical Stokes number                     & \\
   $\v_c$                       & dust collision speed                       & \\
   \hline 
   $\alpha$                   & disk turbulence parameter                             & $\nut / (c_s\,H)$ \\
   $\nut$      & disk turbulent viscosity                             & $\alpha c_s\,H$\\
   $\Sigma_\mathrm{gas}$        & disk surface density                        & $\Sigma_1 \left( \frac{r}{\mathrm{au}} \right)^{-3/2}$ \\
   $T_\mathrm{gas}$             & gas temperature                             & $T_1 \left( \frac{r}{\mathrm{au}} \right)^{-1/2}$ \\
   $c_s$                      & speed of sound                              & $\sqrt{k_B T_\mathrm{gas}/\mu}$ \\
   $r$                         & orbital distance in the disk & \\
   $z$                        & vertical height in the disk & \\
   $H$                        & disk scale height                           & $c_s/\Omega_\mathrm{K}$ \\
   $H_{\rm d}$                        & disk scale height for the dust                            \\
   $\Omega_\mathrm{K}$          & orbital (Keplerian) frequency               & $\sqrt{GM_\star/r^3}$ \\
   $M_\star$                   & stellar mass                                \\
   \hline 
 \end{tabular}
\end{table*}

\begin{acknowledgements}

We thank Satoshi Okuzumi, Zoe Leinhart and Rico Visser for
useful discussions. Most of the simulations 
were done using HPC resources from GENCI-IDRIS (Grant
i2011026174). Part of them were performed on the ``M\'esocentre de
calcul SIGAMM'' at the {\em Observatoire de la C\^ote d'Azur}. 
The research leading to these results has received
funding from the Agence Nationale de la Recherche (Programme Blanc
ANR-12-BS09-011-04).
\end{acknowledgements}

\bibliography{turbulence_and_dust}

\end{document}